\documentclass{emulateapj}
\usepackage{graphicx}
\usepackage{bm}
\newcommand{\Msun}{{\rm M}_{\odot}}
\def\gsim{\mathrel{\rlap{\lower 4pt \hbox{\hskip 1pt $\sim$}}\raise 1pt
\hbox {$>$}}}
\def\lsim{\mathrel{\rlap{\lower 4pt \hbox{\hskip 1pt $\sim$}}\raise 1pt
\hbox {$<$}}}

\newcommand{\eg}{e.g., }
\begin{document}
%%%%%%%%%%%%%%%%%% (1)TITLE PAGE %h
%%%%%%%%%%%%%%%%%
\title{The Mass Spectrum of the First Stars}
\author{Hajime Susa$^1$}
\author{Kenji Hasegawa$^2$}
\author{Nozomu Tominaga$^{1,3}$}
\affil{Department of Physics, Konan University, Okamoto, Kobe, Japan\altaffilmark{1}}
\affil{Center for Computational Science, University of Tsukuba, Japan\altaffilmark{2}}
\affil{ Kavli Institute for the Physics and Mathematics of the Universe (WPI),\\
  The University of Tokyo, 5-1-5 Kashiwanoha, Kashiwa, Chiba 277-8583, Japan\altaffilmark{3}}
\email{susa@konan-u.ac.jp}
%%%%%%%%%%%%%%%%%%%%%%%%%%%%%%%%%%%
% (2)Abstract  & Subject Headings %
%%%%%%%%%%%%%%%%%%%%%%%%%%%%%%%%%%%
\begin{abstract}
We perform cosmological hydrodynamics simulations with non-equilibrium
primordial chemistry to obtain 59 minihalos that host first stars.
The obtained minihalos are used as initial conditions of local three
dimensional radiation hydrodynamics simulations to investigate the
formation of the first stars. 
We find two-thirds of the minihalos host multiple stars, while the
 rest of them have single stars. The mass of the stars found in our
 simulations are in the range of $1M_\odot \la M \la 300 M_\odot$,
 peaking at several$\times 10 M_\odot$.  Most of the very massive stars
 of $\ga 140M_\odot$ are born as single stars, although not all of the single
 stars are very massive. We also find a few stars of $\la 1\Msun$ that
 are kicked by the gravitational three body interactions to the position
 distant from the center of mass. The frequency that a star forming
 minihalo contains a binary system is $\sim 50\%$.
 We also investigate the abundance pattern of
 the stellar remnants by summing up the contributions from the first stars in
 the simulations. Consequently, the pattern is compatible with that of the
 low metallicity Damped Lyman$-\alpha$ systems or the Extremely Metal
 Poor (EMP) stars , if the mass spectrum
 obtained in our experiment is shifted to the low mass side by 0.2 dex.
If we consider the case that an EMP star is born in the remnant of the
individual minihalo without mixing with others, the chemical signature of the pair instability
 supernova is more prominent,
because most of them are born as single stars. 
%Thus the pure abundance pattern of PISN on EMP stars could be found in
% future observations.

\end{abstract}
\section{Introduction}
Formation of the first stars is one of the central objectives of modern
cosmology, which has been intensively investigated from the end of the last
century. We now have the standard model that the first stars
form at the redshift of $z\ga 20$, in the minihalos which are as
massive as $10^5-10^6
M_\odot$\citep{haiman96,tegmark97,nishi_susa99,fuller_couchman00,abel02,bromm02,yoshida03}. 
In such minihalos, the gas temperature
cannot rise up to $10^4$K by gravitational contraction, because the
virial temperature is low. Thus, the only available coolant in the gas
is H$_2$ molecules, since H Ly$\alpha$ cooling is not activated and no heavy elements exist
at that very early Universe.

H$_2$ molecule is an inefficient coolant, because it is a homonuclear diatomic molecule.
Hence, the gas temperature of the primordial gas tend to be higher than that
of the interstellar gas, in which more efficient coolant such as metals
and dusts exist. 
In fact, the gravitationally contracting gas temperature in the
primordial minihalo for $\la 10^{8}{\rm cm}^{-3}$ is  ${200}$K$\sim$ $10^3$K, which is
higher than the temperature of the interstellar gas by roughly two orders of
magnitude. The Jeans mass of the dense core of $\sim 10^4{\rm cm}^{-3}$
in the minihalo is $10^2-10^3M_\odot$, which is much larger than {the present-day counterpart ($\sim 1M_\odot$)}.
In addition, the mass
accretion rate onto the  central star is proportional to $T^{3/2}$, that is
also very high in primordial case. Based upon these theoretical
evidences, the typical mass of the first stars {is} once considered to be very
{high}($\ga 100 M_\odot$).

On the other hand, recent advance {in} the studies on the mass accretion
phase of the first star formation revealed that the mass of the first
stars could be smaller than expected before\citep{clark08,smith11}. The highest
spatial resolution achieved among these studies is less than 1AU,
which enables {them} to resolve the fragmentation of the gas disk that form in the
very vicinity of the primary first star. They found that a ``cluster'' of lower mass stars could
form rather than a single very massive star\citep{clark11a,clark11b,greif11,greif12,machida_doi13}. 
In addition, statistical studies have revealed that there are various
cases of multiplicity in a minihalo, depending on
the properties of the collapsing cloud\citep{stacy13}.

On the other hand, the final mass of the stars are not fixed by the
first several thousand years after the formation of the primary star. It
will take several tens of thousand years until the stars settle
onto the main sequence stars, which is hardly possible to be followed by very
high resolution studies. In addition, ultraviolet radiative feedback
from the protostar comes into play after the mass of the protostar grows
to $\sim 20M_\odot$. Thus, we have to take into account the radiative
feedback properly. 

The effects of ultraviolet radiation from the protostars on the mass
accretion flow were first investigated by \citet{omukai01}. They
constructed a one dimensional spherical steady model of accreting
protostar, and they found the radiation cannot stop the
mass accretion onto the protostar if typical mass accretion rate inferred
by the cosmological simulation is adopted \citep{omukai03}.  
\citet{tanmckee} have constructed a semi-analytical model of the two
dimensional accretion flow, and they found that the mass of the
central protostar grows to at least $\ga 30M_\odot$ to stop the mass
accretion.
 
Following the pioneering works, \citet{hosokawa11} tackled this issue by
a two dimensional Radiation Hydrodynamics (RHD) simulation. As a result,
they found that the mass accretion is shut off by the photoheating of
the ultraviolet radiation from the protostar, when it grows to $43 M_\odot$.
\citet{hirano14} expanded the work to different one hundred cases,
i.e. they picked up 100 minihalos from cosmological simulations, and
followed the formation of the first stars in these halos by two dimensional
RHD simulations. Consequently, they found that the final mass 
spreads in the range of $10M_\odot \la M \la 2000 M_\odot$.

These are two dimensional calculations that can handle the radiation
properly, although it is not possible to investigate the fragmentation
of the gas disk. Thus, we need three dimensional calculations with radiative
transfer in order to consider both {of these} effects.
The three dimensional simulations of this type were first performed by
\citet{stacy12}. They solve the radiation transfer from the primary
protostar, as well as the fragmentation of the disk. As a result, they
found multiple stellar systems and an obvious sign of suppression of mass accretion onto protostars.
However, they have integrated only five thousand years in physical time, which is not
enough to obtain the final mass of the first stars.

In the wake of their works, \citet{susa13} performed a three dimensional RHD simulation, in which
radiation from all protostars are considered. They integrated the
simulation until $10^5$yrs after the formation of the primary star, which
enables them to assess the final mass of the first stars.
They found five stars, in their particular simulation, in the range of
$1M_\odot\la M\la 60M_\odot$, although they significantly underestimate
the radiative feedback effect {because of the insufficient
resolution to capture the propagation of the ionization front}.
This mass range is, however, {just} one result from one realization. As pointed
out by \citet{stacy13} and \citet{hirano14} we need a statistical study to obtain
the {correct} mass range of the first stars, starting from cosmological simulations.

The mass range of the first stars is also important to clarify
 chemical evolution of the early Universe because nucleosynthesis in the
 first stars and their supernovae, which provides the first metal
 enrichment in the Universe, depends on their masses. The chemical
 evolution of the early Universe and the mass range of the first stars
 have been observationally constrained so far by means of
 observations of elemental abundances of metal-poor stars that formed in
 the early Universe \citep[e.g.,][for a review]{bc05}. In combination
 with theoretical studies of supernova nucleosynthesis and chemical
 evolution \citep[e.g.,][for a review]{nom13}, the following
 consequences {have been} obtained: no chemical signature of a pair-instability
 supernova (PISN) yielding a peculiar abundance pattern has been found in
 the metal-poor stars \citep{heg02,ume02} and the
 abundance patterns of the metal-poor stars are well reproduced by
 supernova explosions of first stars as massive as supernova progenitor
 stars in the present day ($M\la 100M_\odot$) \citep{tom07b,heg10}. 
 %On
 %the other hand, it has not been available to theoretically predict the
 %elemental abundance of the early Universe based on cosmological
 %simulations. 
 On the other hand, theoretical predictions based on cosmological simulations
   have not been available for the elemental abundance of the early Universe.
 This is because
 the masses, i.e., initial mass function (IMF), of the first stars have
 not been well constrained.

In this paper, we perform cosmological simulations to obtain 59
minihalos that will host first stars. These halos are used as the initial
conditions {for} long term RHD simulations {of} the first star formation. Based upon the
results of the simulations, we derive the range of the stellar mass and the correlation
between the stellar mass and the properties of the host minihalos/clouds.
We also discuss the properties of the formed stellar systems in the
minihalos, as well as their chemical imprints on the metal poor systems.
This paper is organized as follows. We describe the method and setup of
the numerical experiment in sections 2 and 3. Then we show the results
of the simulations in section 4. Section 5 is devoted to discussion, and
we summarize in section 6.

\section{Cosmological simulation}
\label{c_simulation}
In order to collect the samples of the minihalos that host the first
stars, we perform cosmological hydrodynamic simulations using START\citep{hasegawa10} that is
an RHD code used in \cite{HS13}. We solve not only hydrodynamics but
also non-equilibrium chemistry regarding $\rm H, H^{+}, e^{-}, H_2^+,
H^-, H_2, He, He^+$, and $\rm He^{2+}$. The code can handle the
ultraviolet radiation from various sources\citep{hasegawa09,umemura12}.
However, in this work, we do not take into account the external
radiation sources, in order to concentrate on the formation process 
of the so-called population III.1 stars.
We will explore the impact of the external radiation sources on the
minihalos and the mass spectrum of the first stars, i.e. the population
III.2 stars in a forthcoming paper. 

We employ a simulation box of 100kpc (comoving) on a side, which contains $2\times512^3$ SPH and DM particles. 
The initial conditions are generated by 2LPT code, {in which the second-order Lagrangian perturbation theory is used,} \citep{Crocce06} and all the simulations are initialized at $z=200$. 
Throughout this paper we assume the $\Lambda$ cold dark matter Universe with cosmological parameters of $\Omega_b = 0.049$, $\Omega_0=0.27$, $\Omega_\Lambda=0.73$, and $h=0.71$, based on 7-year WMAP results \citep{Komatsu11, Jarosik11}. 
We also adopt artificially enhanced normalizations of the power spectrum, $\sigma_8=1.5$, and $1.2$ to accelerate the structure formation in the same manner as previous studies \citep[e.g.][]{stacy12}. 
Hereafter, we  call the former series of runs  Runs A and the
latter Runs B. 
We also perform a reference cosmological simulation with 2Mpc box
containing $2\times 1024^3$ SPH and DM particles, to check if the
properties of the minihalos in Runs A/B are
similar to those with a normal $\sigma_8=0.82$ in larger box.  

In cosmological hydrodynamic simulations, we often suffer from the fact
that a time step becomes very short during a density peak is
collapsing, because of the very short thermal time scale at the peak. 
Consequently, it is difficult to extract several
samples of minihalos from a cosmological run. 
In order to avoid such a difficulty, we employ a procedure described below. 
In each run, we adopt a threshold density $n_{\rm th}$. When a density
peak reaches $n_{\rm th}$, we extract and store the data of all the physical
quantities around the peak, i.e. the data of a minihalo.
Then we convert all of the SPH particles that
reside within 150pc from the density peak, into collisionless particles,
retaining the information of the positions and velocities of the
particles. Thus, the very short thermal time scale around the density
peak is safely ignored that allow the other minihalos to collapse.
{We also note that the halos composed of 
the collisionless particles could merge with neighboring non-collapsed
halos, although such events are very rare. 
We never extract halos which are contaminated by the  
the collisionless particles when the halos collapse. 
In fact, we find only five cases of such merger event in all of the Runs A/B. 
Hence, neglecting the contaminated halos hardly affects our results.}

Owing to this procedure, we can alleviate the difficulty caused by
shortened time steps, and extract multiple minihalos from each run. We
set $n_{\rm th}=10^8 {\rm cm^{-3}}$ for both of the 
Runs A and B, and $n_{\rm th}=10^3 {\rm cm^{-3}}$ for the
reference run. 

We note that the minihalos found in the reference run are never used for the subsequent RHD simulations (see section \ref{RHD} for more details), since the DM potential well seems to be still non-negligible when the density peaks reach $n_{\rm th}=10^3 {\rm cm^{-3}}$ 
{\citep{abel02}}. 
We carry out 24 runs of Runs A and 17 runs of Runs B, and extract 59 halos from the runs. 
The characteristics of the cosmological simulations are summarized in Table \ref{cosmo}. 

\begin{table*}
\caption{Characteristics of cosmological runs}
\begin{center}
\begin{tabular}{cccccccc}
  \hline
  Series  
  & $L_{\rm box}$ [Mpc] %\footnote{The length of simulations box.}
  & $N_{\rm SPH}$ (=$N_{\rm DM}$) %\footnote{The number of SPH (DM) particles.}
  & $m_{\rm DM}$ $[M_{\odot}]$ %\footnote{The mass of a SPH particle.}
  & $m_{\rm SPH}$ $[M_{\odot}]$  %\footnote{The mass of a DM particle.}
  & $\sigma_8$ & $N_{\rm halos}$ %\footnote{The normalization of the power spectrum.}
  & $n_{\rm th}{\rm [cm^{-3}]}$ %\footnote{The threshold density (see text for details). } 
  \\
  \hline
   Runs A & 0.1 & $512^3$ & $0.25 $  & $0.049 $ &  1.50 & 38 & $10^8$ \\
   Runs B & 0.1 & $512^3$ & $0.25 $  & $0.049 $ &  1.20 & 24 & $10^8$ \\
   Reference & 2.0 & $1024^3$ & $250 $  & $4.90 $ &  0.82 & 1878 & $10^3$  \\
   \hline
\end{tabular}
\end{center}
\label{cosmo}
\end{table*}
\section{The local simulation of first star formation}
\label{RHD}
\subsection{RHD simulation}
We pick up 59 minihalos from the cosmological simulations, then follow the
subsequent evolution by local RHD simulations. 
Each SPH particle in the minihalos is split into ten SPH particles in order to achieve
sufficient resolution to resolve the fragmentation at $100-1000$AU scales.
As a result, the mass of an SPH particle in the local simulations is
$m_{\rm SPH}=5\times 10^{-3}M_\odot$ which corresponds to the mass resolution of
$0.5M_\odot (=2N_{\rm neib}m_{\rm SPH})$, where $N_{\rm neib}(=50)$ is the
number of neighbor particles\citep{bate_burkert97}. We perform local simulations starting from
these initial conditions.

The code we use to simulate the systems 
is RSPH \citep{susa04,susa06,SU06} with some extensions (sink particles,
updated rates, cooling rates at high density, time stepping) described in \citet{susa13}.
Utilizing the code, we solve the hydrodynamics, non-equilibrium
primordial chemistry of
six species, e$^-$, H$^+$, H, H$_2$, H$^-$, H$^+_2$, and radiative
transfer of ultraviolet photons in this numerical experiment. 
RSPH is a code that solve hydrodynamics by SPH scheme, and the radiation
transfer by ray-tracing. The ray-tracing from a point source is realized
by connecting neighbor particles using the neighbor list of SPH method\citep{susa06}.
The ray-tracing enables us to calculate the optical
depth at the Lyman limit as well as the H$_2$ column density from any
source star to each SPH particle. {We remark that the
resolution of the present simulations is not enough to capture the
propagation of the ionization front, although the
photoionization/photoheating is implemented to the code (see 4.1 and 5.1).}
%We can create the tables of the photoionization/photoheating rates as
%functions of optical depth by integrating the
%spectrum before we start the simulation\citep[e.g.][]{susa04}.
H$_2$ column density is also used to calculate the self-shielding function of
Lyman-Werner (LW) photons\citep{wolcott-green11}.
H$_2^+$ photodissociation is also taken into account, based upon the cross-section in
\cite{stancil94}. H$^-$ radiative detachment is assessed using the
fitting formula for the cross-section in \citet{tegmark97}. 

We take into consideration the standard cooling processes of primordial gas such as H/H$_2$ line
cooling, H$_2$ formation heating/dissociation cooling, H
ionization/recombination cooling, bremsstrahlung, optically thin H$^-$
cooling and collision induced emission cooling.
%The self-shielding effect of H$_2$ bound-bound emission is taken into
%account by the shielding function\citep{ripamonti}.
{Optically thick H$_2$ line cooling is taken into consideration
by using the simple analytic formula shown as the equation (22) in \cite{ripamonti}}.
In the present version of RSPH code based upon \citet{susa13}, we
update the hydrodynamics, gravity, and radiative transfer at the time
step given by the Courant condition, while the energy
equation and chemical reaction equations are integrated at smaller time steps.

{We remark that we only simulate the central spherical region with radius
0.6pc in the cloud, after the formation of the first sink in
order to save the computational time.  The outer envelope of $r > 0.6$pc
is omitted from the calculation since it hardly affects the evolution of
the inner region within $10^5$yrs.}

\subsection{Sink particles}
%%how to create 
A sink particle technique is employed in the present simulations, which
is {the} same as the one employed in \citet{susa13}.
%We put a density threshold of $n_{\rm
%sink}=3\times10^{13}{\rm cm^{-3}}$ above which the SPH particle is changed into a sink particle.
We search the highest density peak at every time step. Then
corresponding SPH particle is converted into a sink particle if the peak
density exceeds a threshold density, $n_{\rm sink}=3\times10^{13}{\rm cm^{-3}}$.
We also set an accretion radius as $r_{\rm acc}=30{\rm AU}$. If the
distance from a sink particle to an SPH particles is less than $r_{\rm
acc}$ and they are gravitationally bound with each other, the SPH
particle is  merged to the sink particle, conserving the linear momentum
and the mass. These sink particles are regarded as protostars in this experiment.

%% sink-sink merging omitted 
The sink-sink merging is not allowed in the present numerical
experiment, because the actual protostellar radius is much smaller than
$r_{\rm acc}$. In fact, the radius of the protostar is less than $\sim
1$AU\citep{hosokawa09} even at the maximally expanded phase just before
Kelvin-Helmholtz(KH) contraction. 
However, we have to keep in mind that merging between the protostars should be properly taken into consideration at much higher resolution studies\citep[e.g.][]{greif12,machida_doi13}.
 
%% sinks are like BH, actual mass is smaller
We also assume that the sink particles behave as ``black holes'',
i.e. they interact with  surrounding particles as only sources of
gravity (and radiation, discussed later). 
In other words, 
the pressure forces from sink particles to surrounding SPH particles are
omitted. The recipe of the sink particles
employed in the present work is known to overestimate the mass accretion
rate\citep[e.g.][]{bate95,bromm02,martel06}. Combined with the fact that
the accretion radius is much larger than the protostellar radius,
the resultant mass of the formed sink particles would be larger than the
actual mass of the first stars.

We stop the simulation at $10^5$yrs after the first sink formation,
which is longer integration time than the two dimensional RHD simulation by \citet{hosokawa11}.

\subsection{Protostellar evolution model}
We turn on the sinks, i.e. protostars,  when they are created.
In order to obtain the luminosity/spectrum of the source protostars found in
the simulations, we have to employ a protostellar evolution model.
These properties of a protostar is
obtained based on the calculation by \citet{hosokawa09}.
They have calculated the {steady} evolution of the protostars\citep{SPS1986} with given (fixed)
mass accretion rates. We use their numerical data of $\dot{M}_* = 4\times
10^{-3}, 10^{-3}, 10^{-4} \,{\rm and }\, 10^{-5} M_\odot
{\rm yr}^{-1}$, and analytic formula for $\dot{M}_* > 0.1 M_\odot {\rm
yr}^{-1}$. Intermediate cases are interpolated between the data.
The analytic formula for high mass accretion rate is given by
\citet{hosokawa12}.
{The analytic formulae are given in the two phases, i.e. the mass
accretion phase and the  KH contraction phase. The protostars normally
evolve from the former to the latter, and the phase changes when the
mass of the protostar exceeds $M_{*,{\rm teq}}$. This corresponds
to the time when the mass accretion time scale, $t_{\rm acc}$
equals to the KH contraction time scale, $t_{\rm KH}$
\citep[e.g.][]{hosokawa12}. These quantities are defined as:}
\begin{eqnarray}
 M_{*,{\rm teq}}&=&14.9M_\odot\left(\frac{\dot{M}_*}{10^{-2}M_\odot {\rm
			       yr}^{-1}}\right)^{0.26}\nonumber\\
t_{\rm acc}&=& \frac{M_*}{\dot{M}_*}\nonumber\\
t_{\rm cool}&=& \frac{GM_*^2}{R_*L_*}\nonumber
\end{eqnarray}
where $R_*, L_*, M_*$ and $\dot{M}_*$ denote the radius, the luminosity,
the mass and the mass accretion rate of the protostar, respectively.

For the mass accretion phase($M_* < M_{*,{\rm teq}}$) the analytic formula
reads, 
\begin{eqnarray}
R_*&=& 26
 R_\odot\left(\frac{M_*}{M_\odot}\right)^{0.27}\left(\frac{\dot{M}_*}{10^{-3}M_\odot{\rm
	 yr}^{-1}}\right)^{0.41}\nonumber\\
L_*&=& \frac{GM_*\dot{M}_*}{R_*}\nonumber\
\end{eqnarray}

\begin{figure}
\plotone{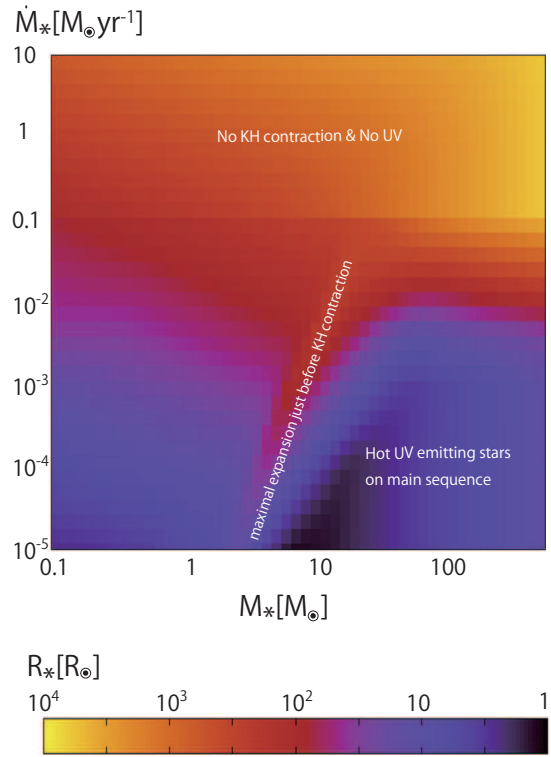}
\caption{Color contour of protostellar radius on $M_* - \dot{M}_*$ plane.
}
\label{fig_protostar}
\end{figure}

In KH contraction phase ($M > M_{*,{\rm teq}}$), we have
\begin{eqnarray}
R_*&=&2.6\times
 10^3R_\odot\left(\frac{M_*}{100_\odot}\right)^{1/2}\nonumber\\
L_*&=& 3.8\times 10^6 L_\odot\left(\frac{M_*}{100_\odot}\right).\nonumber
\end{eqnarray}
Fig.\ref{fig_protostar} shows the color contour of $R_*$ as a function
of $M_*$ and $\dot{M}_*$.
The lower right black-purple region corresponds to the parameter space
where the protostar is settled onto the main sequence stars emitting UV radiation, whereas the
top orange-yellow region is no UV emitting high mass accretion branch
found by \citet{hosokawa12}. The ridge line found in the lower middle
region denotes the maximal expansion phase of the protostar just before
the KH contraction.
We assume the spectrum of the protostar is black body, and the effective temperature of the protostar is obtained by the equation
$L_* = 4\pi R_* \sigma T_{\rm eff}^4$. 
We also have similar contour map of $L_*$, thereby we can derive the luminosity
and spectrum of a protostar with given $M_*$ and $\dot{M}_*$
\footnote{We have to keep in mind the limitation of this procedure. In case the
mass accretion is very {stochastic} and changes much faster than the KH
contraction time scale, this procedure is not sufficient to trace the
evolution of the protostar. In the present simulations, the mass
accretion rate is smoothed over $10^3$yrs, which is longer than the KH
contraction time scale during the UV emitting
phase of protostars. Thus, the present treatment is self-consistent.}.

On the other hand, we can assess the protostellar mass ($M_*$) and the
mass accretion rate ($\dot{M}_*$) self-consistently from the
hydrodynamics simulation. 
The mass accretion rates onto the sinks in the present simulation are obtained by
averaging over $10^3$yrs in order to avoid artificial jumps due to
SPH discreteness. 
These quantities are fed to the protostellar
evolution model described above at every time step, which in turn gives
the luminosity and spectrum of protostars used in the simulation at the next step.
Hence, the protostellar evolution model is self-consistently taken into
account in the radiation hydrodynamics calculations.

\section{Results}
\subsection{Properties of the minihalos}
\label{results_cosmological}
\begin{figure}
	\centering
	{\includegraphics[width=8cm]{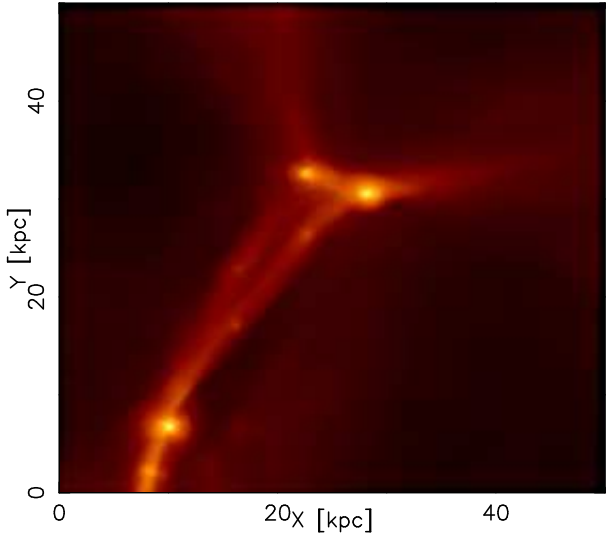}}	
%	\caption{Projected gas density distribution in one of Runs A at $z=14.0$. 
%	The three density peaks in the map respectively correspond to Halo 001, 002, and 003 in Fig \ref{Hhalos_zoom}.}
	\caption{Projected gas density distribution in one of Runs A at
 $z=14.0$. {The presented volume is 1/8 of the whole box of this
 run.}  
The three density peaks in the map are extracted as the hosts of
 the first stars.}
	\label{cosmicweb}
\end{figure}
%\begin{figure}
%	\centering
%	{\includegraphics[width=8cm]{Halos_sigma1.5G.eps}}	
%	\caption{Projected gas density distributions of all halos in Runs A. The size of the area shown in each panel is 100pc (physical)
%	on a side. The gray scale contour indicates gas mass density.}
%	\label{Hhalos_zoom}
%\end{figure}
%\begin{figure}
%	\centering
%	{\includegraphics[width=8cm]{Halos_sigma1.2G.eps}}	
%	\caption{Same as Fig.\ref{Hhalos_zoom}, but for Runs B.}
%	\label{Lhalos_zoom}
%\end{figure}
\begin{figure*}
	\centering
	{\includegraphics[width=12cm]{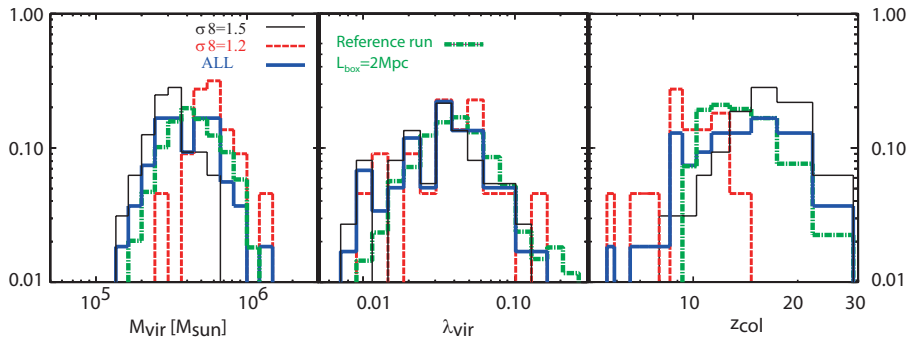}}
	\caption{Frequency distributions of the virial mass, spin parameter, and collapse redshift of halos. 
	In each panel, the result of Runs A, and Runs B are respectively indicated by 
	the {thin} solid and dashed lines. Also, the results for all halos,
 namely combined data of Runs A and Runs B, 
	are shown by the {thick solid} lines. The results of the reference run are indicated by the dot-dashed lines. } 
	\label{his_halo}
\end{figure*}

\begin{figure}
	\centering
	{\includegraphics[width=6cm,angle=270]{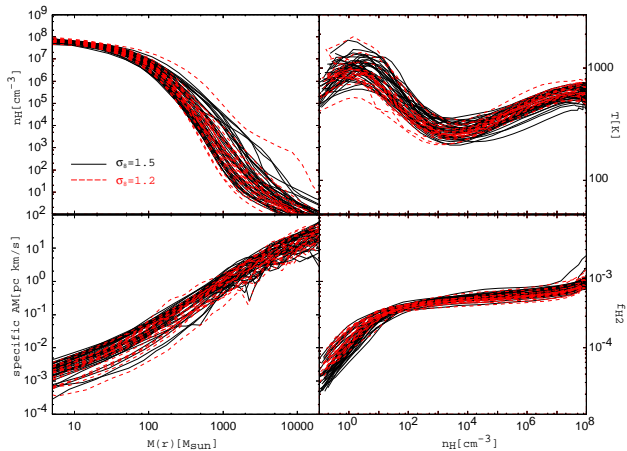}}	
	\caption{Number density as a function of enclosed mass (upper left), specific angular momentum as a function of enclosed mass 
	(lower left), 
	gas temperature as a function of number density (upper right), and $\rm H_2$ fraction as a function of number density (lower right). 
	In each panel, the Runs A and Runs B halos are indicated by the solid and dashed curves, respectively. }
	\label{prop_halo}
\end{figure}
\begin{figure*}
	\centering
	{\includegraphics[width=12cm]{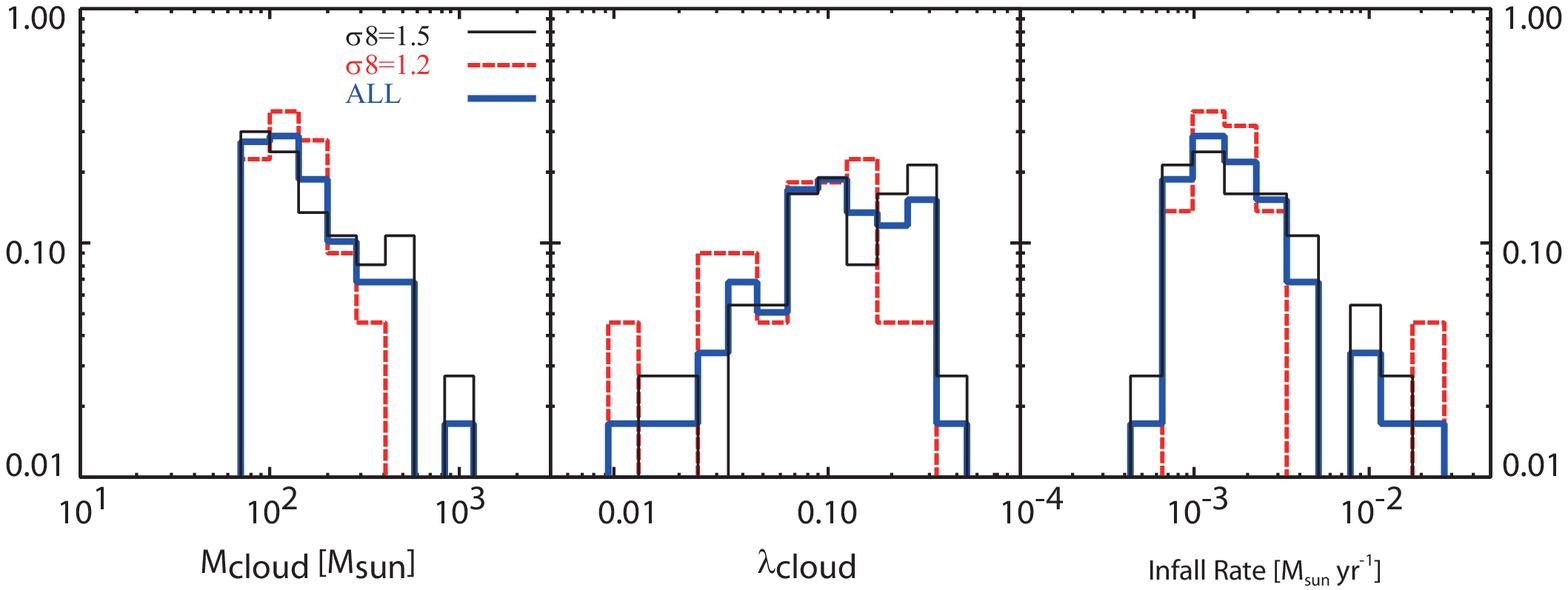}}	
	\caption{Frequency distributions of the mass, spin parameter, and infall rate of clouds. 
	In each panel, the result of the Runs A, and Runs B are respectively indicated by 
	the {thin} solid and dashed lines. Also, the results for all halos,
 namely the combined data of Runs A and Runs B, 
	are shown by {thick solid} lines. }
	\label{his_cloud}
\end{figure*}

In this section, we describe the properties of the minihalos found in
the cosmological simulations. 
%we firstly confirm whether or not the properties of minihalos are affected by the artificially enhanced $\sigma_8$. 
In order to quantify the properties of the dark matter halos, we define
the region, whose mean mass density inside a certain radius from a
density peak corresponds to 200 times the average mass density of the Universe, as a virialized halo. 
Then we define the mass inside the radius, called virial radius $r_{\rm vir}$, as the virial mass $M_{\rm vir}$. 
In addition, when the density peak of a halo reaches $n_{\rm th}$, we regard the halo as collapsed. 
We estimate the spin parameter of a halo using the following formula,  
\begin{equation}\label{spin}
	\lambda_{\rm vir}= \frac{j_{\rm vir}}{\sqrt{2GM_{\rm vir}r_{\rm vir}}}, 
\end{equation}
where $j_{\rm vir}$ is the specific angular momentum of the halo. 
 
In Fig.\ref{cosmicweb}, we show the {zoomed} map of the projected gas density at $z=14$
from a run with $\sigma_8=1.5$ (Runs A). In this case, we find three
minihalos (three bright spots in the map) that host first stars. 
%We perform 24 runs of Runs A and 17 runs of Runs B, and extract 59 halos.

In Fig. \ref{his_halo}, we show the frequency distributions of the halo virial mass, the spin parameter, and the collapse redshift obtained by our simulations . 
In each panel of the figure, the heights of the histogram is normalized
by the total number of the halos. 
It is needless to say, the halos in Runs A tend to collapse earlier than
those in  Runs B because of the difference of the mean heights of density peaks. 
 We also can see that the halos in Runs A are relatively less massive than those in Runs B. 
This trend also can be understood by the fact that the first star
forming halos have to be cooled by the H$_2$ cooling process, 
which can only be important 
if the temperature of the collapsing gas clouds exceed $1000-2000$K. 
Since the maximal gas temperature is basically determined by
the virial temperature of the halos, {which is proportional to $M_{\rm vir}^{2/3}(1+z_{\rm col}$)}, 
the star forming minihalos collapse at lower
redshift are more massive to meet the condition of virial temperature.

%This trend results from the higher virial mass with a given virial
%temperature for  halos in Runs B. 

We also notice that the spin parameters for the halos in
Runs B is slightly higher than that in Runs A. This
seems to originate in the fact that higher density perturbations reach
their turn around time earlier, thereby their spin-up time tends to be
shorter \citep{SB95}. As argued in the above, our samples of minihalos are
qualitatively consistent with the theoretical predictions.
 
%In addition, the properties of halos calculated with the enhanced $\sigma_8$ are well consistent with those with the normal $\sigma_8$. 
%Therefore we conclude that our samples likely be regarded as normal
%virtualized halo (?}???????÷????????????reference run??????????????????
%???????\??????).
We also tested if the properties of the minihalos in Runs A/B are similar to
those found in the reference run. The volume in the reference run is
8000 times larger than that of Runs A/B, and the employed $\sigma_8$ is
0.82 which is the ordinary value in WMAP 7-year cosmology. Thus, the
reference run would give {proper} average distribution of minihalos,
although it lacks the mass resolution required {for} the
initial conditions of the local RHD simulations. We note that the
reference run is stopped at $z=9$ to save the computational time.
We pick up 1878 minihalos
that will host first stars in the reference run, and the results are superimposed
on the panels of Fig.\ref{his_halo}. 
Consequently, we find that the frequency distribution of the minihalos
picked up from the boosted $\sigma_8$ runs (Runs A + Runs B, blue dotted
line ``ALL'') are consistent with those with the normal $\sigma_8$ for $z > 9$.
The chief reason would be that the first star forming minihalos form at
the high-$\sigma$ density peaks even in the larger box, which makes the properties of the minihalos in the reference run to
be similar to those in Runs A/B.

In Fig. \ref{prop_halo}, we show the number density and specific angular
momentum as a function of enclosed mass, and the gas temperature and
hydrogen molecular fraction as a function of number density. Here, we
emphasize two important points in this figure. First, these internal
properties of our samples are similar to those in previous studies
\citep{yoshida06, ON07, hirano14}. This also indicates that our samples
are hardly affected by the enhanced normalization of the density fluctuations.  
Second, the properties of the halos shown in the figure show large variations. These variations lead to the variety of clouds.   

\subsection{Properties of the cloud}
{We also define the dense ``clouds'', the
hosts of the first stars,  located at the center of the gravitational potential of
the minihalos, and show the properties of them.}
The definition of {a} "cloud" is the same as that adopted in
\cite{hirano14}. 
We {take the snapshot when the peak density reaches $n_{\rm
th}=10^8{\rm cm^{-3}}$, and} average the physical quantities over the thin spherical shells, to
obtain the radial distribution of them.
Then we define the radius where the ratio of the enclosed
mass to the local Bonner-Ebert (BE) mass takes the maximum as the cloud
radius $r_{\rm cloud}$ 
\footnote{We have to keep in mind
that this definition could
introduce artificially large cloud mass because it can pick up
relatively distant radius when multiple density peaks are present.}.
Here the local BE mass is given by 
\begin{eqnarray}
	M_{\rm BE} &= & 1050 \left( \frac{T}{200{\rm K}}\right)^{1.5} \left( \frac{\mu}{1.22}\right) ^{-2.0} \nonumber\\
	& & \left( \frac{n_{\rm H}}{10^4{\rm cm^{-3}}}\right) ^{-0.5}\left(\frac{\gamma}{1.66}\right)^{2.0}M_{\odot}, 
\end{eqnarray}
where $\mu$ and $\gamma$ are the mean molecular weight and the adiabatic index, respectively. 
Once we determine the cloud radius $r_{\rm cloud}$, the cloud mass is defined as the enclosed mass within $r_{\rm cloud}$. 
The cloud radius ($r_{\rm cloud}$) roughly corresponds to the radius
where $n_{\rm H}\simeq 10^4{\rm cm^{-3}}$ is satisfied. 
The spin parameter of a cloud is estimated by the equation(\ref{spin}), replacing the subscript of "vir" with "cloud". 
The mass infall rate at a cloud surface is defined as 
\begin{equation}
	\dot{M}(r_{\rm cloud}) = 4\pi r_{\rm cloud}^2 \rho(r_{\rm cloud}) v_{\rm rad}(r_{\rm cloud}), \label{eq:infall}
\end{equation}
where $v_{\rm rad}(r)$ and $\rho(r)$ are the radial velocity and gas mass
density averaged over a thin shell at a radius $r$, respectively. 

We show the frequency distributions of the cloud mass, the spin parameter, and the infall rate at the surface of the cloud in Fig. \ref{his_cloud}. 
As expected from Fig. \ref{prop_halo}, we can see large variations in
the properties of the clouds. 
Also, it is important to remark that the fraction of less massive cloud with $M_{\rm cloud}<100M_{\odot}$ is smaller than that in \cite{hirano14}. 
{They found $\sim 1/3$ of the minihalos host such less massive
clouds}. It likely originates in the lack of HD cooling in our simulations. As pointed out by \cite{hirano14}, HD cooling would be important for our sampled halos that cool down to $T\approx 200 \rm K$.  

{The mean spin parameter of clouds in Runs A is slightly higher than
that in Runs B, in contrast to the fact that mean spin parameter of
halos in Runs A is slightly lower than that in Runs B. It is not simple
to give a comprehensible reason for the trend, since the angular
momentum of the central baryonic component is determined at much smaller
scale than the halo scale. }

\subsection{An example among the local simulations}
We obtain 59 minihalos that are going to host first stars from the
cosmological simulations as described above. The next step is to perform local
RHD simulations of first star formation using these minihalos as the initial
conditions.
In this section, we show the results of a particular case among these
minihalos {in which the low mass stars form as did in the case shown
in \citet{susa13}}. The present result is consistent with that of \citet{susa13}.

\begin{center}
\begin{figure}
\plotone{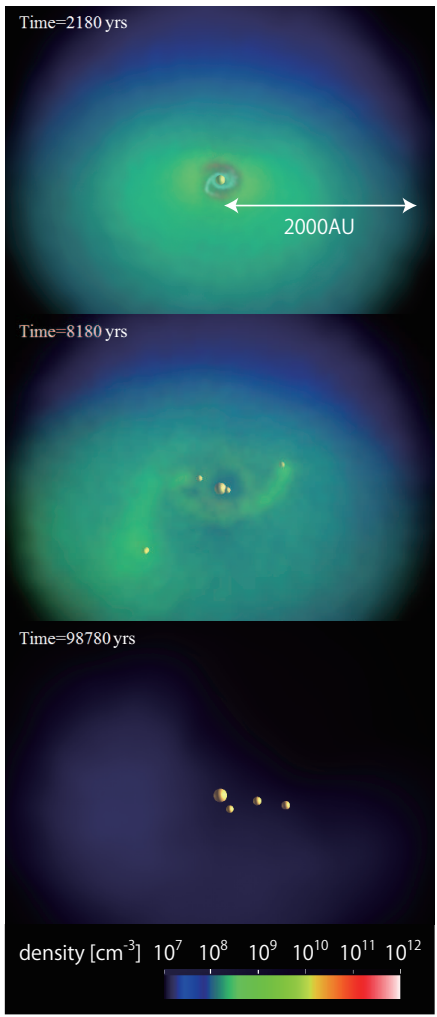}
\caption{Density distributions at three snapshots, 2180 yrs,
 8180 yrs  and 98780 yrs after the formation of the primary star by
 pseudo volume rendering. White spheres represent the sink particles, and
 the size is proportional to their mass.}
\label{fig_VR}
\end{figure}
\end{center}
\begin{figure}
\plotone{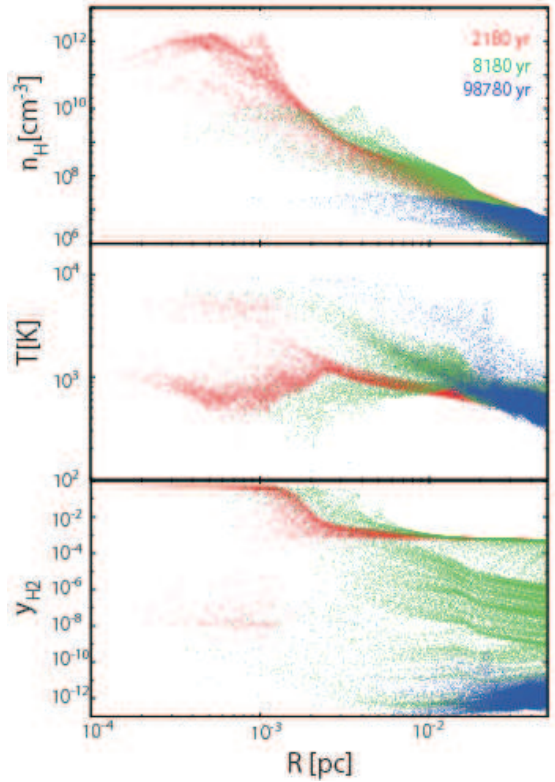}
\caption{Density (top), temperature(middle) and H$_2$ fraction(bottom)
 are shown as functions of the distance from the primary star for a
 typical minihalo. Each dot corresponds to the SPH particle. Three colors
 corresponds to three snapshots at 2180 yrs, 8180 yrs  and 98780 yrs after the formation of the primary star.}
\label{fig_R-X}
\end{figure}

In Fig.\ref{fig_VR}, we show the density distribution of the gas within
2000AU {($=10^{-2}$pc)}in radius around the primary first star at three epochs (hereafter we call the most massive star at the final phase primary star).  The
color denotes the density, while the small spheres represent the
positions of the sink particles, and their radii are proportional to the
mass of the sinks. In the early phase of the mass accretion (2180yr
after the first sink, top), dense accretion disk form around the
primary star, and we find  a prominent spiral structure as was also found
in previous works \citep[e.g.][]{susa13}. After a while, the spiral arms
fragment into sink particles, and the gas density surrounding the sink
particles decline because of the radiative feedback from the stars
(8180yr, middle). Finally, the gas density around the sink particles
becomes much lower than the initial disk, that will limit the further
mass growth of the sinks (98780yr, bottom).
{The masses of the sinks at the final stage of the simulation are in
the range of $4 M_\odot\la M_* \la 40 M_\odot$.}

Fig.\ref{fig_R-X} shows the number density of the
gas $n_{\rm H}$ (top), the gas temperature (middle), and the H$_2$ fraction $y_{\rm H_2}$ (bottom) as
functions of the distance from the primary star. Each dot corresponds to
an SPH particle, and the three colors represent three snapshots at
different epochs which are equivalent to those in Fig.\ref{fig_VR}. 

In the early phase, 2180yrs after the formation of
 the primary star, an accretion disk of $n_{\rm H}\simeq 10^{12}{\rm cm^{-3}}$ forms at inner
{$10^{-3}$pc ($\simeq 200$AU)} region (top, red dots). The temperature of the disk is
$\la 1000$K(middle, red dots) and it is fully molecular(bottom, red dots).
We also observe less dense
particles of $10^{10}-10^{11}{\rm cm^{-3}}$ at {$r < 10^{-3}$pc} (top, red dots), these are
located on the polar region of the primary star. 
The temperature of these gas particles are as {low} as several $\times 10^3$K, and H$_2$ molecules are dissociated.
The gas is heated by the chemical heating of H$_2$ formation
 process which is prominent because of the presence of the
 photodissociative radiation from the protostar\citep[][see also section
 \ref{discussion}]{susa13}.
{We also note that \citet{turk10} reported the importance of chemical heating even before the protostar formation.}
As the time proceeds, the gas density around the central protostar is
getting lower and lower (top, green/blue dots), and the temperature is
kept around $10^3{\rm K} \la T \la 10^4{\rm K}$ (middle , green/blue dots). 
At the final stage, the gas is totally dissociated and the density is as
high as $10^7{\rm cm^{-3}}$. 

\begin{figure}
\plotone{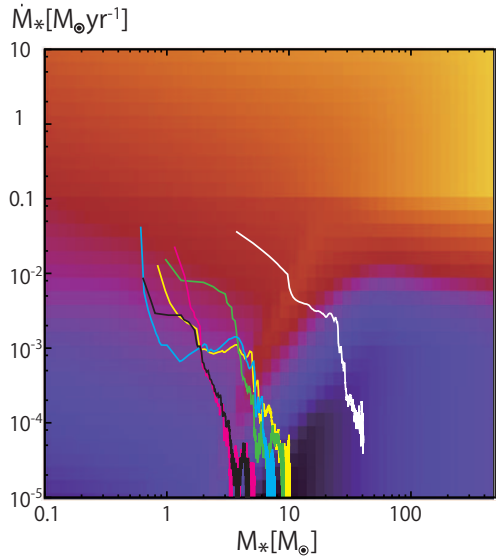}
\caption{Evolution of sink particles on the $M-\dot{M}$ plane. The
 color gradient is same as Fig.\ref{fig_protostar}. Solid lines
 denote the path of the six sink particles in this particular run.
}
\label{fig_mmdot}
\end{figure}

Fig.\ref{fig_mmdot} shows the evolution of the six sink particles born in this
particular case on the $M_*-\dot{M}_*$ plane. 
The final mass of the primary one is as large as $40M_\odot$, while the
others are $\la 10M_\odot$. The final mass accretion rates are $\dot{M}\la
{\rm a~few} \times 10^{-5}M_\odot~{\rm yr}^{-1}$, which is much lower than the initial
rate of $10^{-2}-0.1 M_\odot{\rm yr}^{-1}$ because of the radiative
feedback effects. 
%$10^{-5}M_\odot{\rm yr}^{-1}$ means it can gain only 10$M_\odot$ after $1$Myrs integration.
In addition, the mass accretion rate at such later phase would be
lower than this value, if we could take into consideration the
photoionization feedback properly\citep{hosokawa11}, since the ionized
gas has higher temperature. 
%We will discuss the difficulty of resolving
%the ionization front in section \ref{discussion}. 

We also remark that the primary and the secondary have already settled down to
the main sequence phase at $10^5$yr. The rest of the low mass stars are uncertain,
since it is not possible to resolve the mass accretion rate
$\la 10^{-5}M_\odot$yr$^{-1}$, above which the stars of $\la 10M_\odot$
are still in the pre-main-sequence phase.
We also have checked that all of the protostars more massive than
$10M_\odot$ found in 59 minihalos are in the main sequence phase by
the end of the simulation, i.e. $10^5$ yrs after the first sink formation.

%This is due to the strong
%radiative feedback effect already found in previous works \citep{hosokawa11,stacy12,susa13}.

\subsection{Mass spectrum}
\label{IMF}
\begin{figure}
\plotone{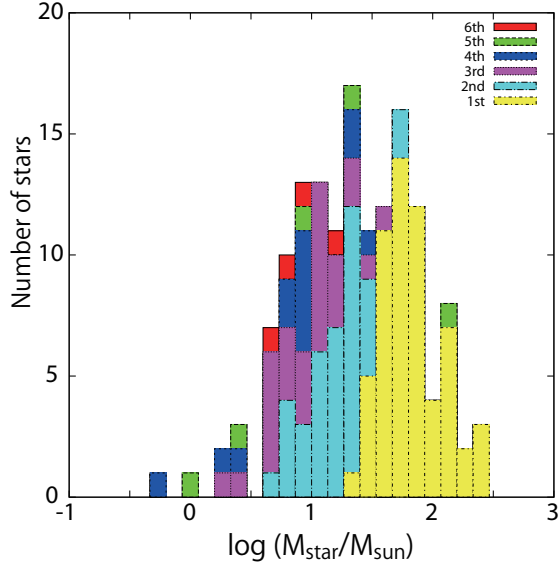}
\caption{The mass spectrum of first stars is shown. Colors in the
 histogram correspond the order of birth of these stars. The color
 legend in the upper right corner describes the correspondence between
 the order of birth and the color.}
\label{fig_spectrum_all}
\end{figure}

We perform local radiation hydrodynamics simulations starting from the
59 minihalos found in the cosmological simulations.
Hence we obtain the mass spectrum of the stars by summing up the
contributions from all the minihalos.
In the mass spectrum of Fig. \ref{fig_spectrum_all}, all the stars found in the local
simulations are taken into account.

It is immediately obvious that we
have a very top heavy mass spectrum with a peak at
several tens of solar mass, and most of the first stars are within the
range of $10M\odot \la M \la 100M_\odot$. 
This is the first IMF of the first stars {by way} of the three dimensional simulations including the effects of
the radiative feedback and the fragmentation.

On the other hand, stars exceeding $140 M_\odot$, i.e. the progenitors of
PISNe also exist in the
simulations.
In fact, those very massive stars account for $\sim$20\% of the total
mass accreted onto the stars found in the simulations.
However, as will be discussed in section \ref{discussion}, the effects of
the radiative feedback tend to be underestimated in the present
simulations. 
Thus, the high mass
end of the present mass spectrum could be larger than the real
spectrum. Thus, number of stars of exceeding $140 M_\odot$ presumably smaller
than the present results, and the mass fraction of $20\%$ should be
regarded as an upper limit.

We also mention that we do not find the case where very massive stars of
$>300M_\odot$ form {unlike those} found by \citet{hirano14}. 
The branch to form such stars requires
very high mass accretion rate of $\ga 10^{-2}M_\odot {\rm yr}^{-1}$ all
through the growth of the protostar. This short mass accretion time
scale does not allow the KH contraction of the protostar, which
inhibits
the ultraviolet radiative feedback. In the present calculations,
the mass accretion rate onto each star does not stay at such
{a} high {value}.
One possible reason is the fragmentation of the accretion disk, 
{which} simply reduces the mass of the primary stars. 
In addition, the mass accretion rate onto a star {is also
reduced}, which makes the mass accretion time scale longer. As a result,
the radius of the protostar shrinks to emit ultraviolet radiation, 
and the radiative feedback reduces the mass accretion rate.
As for the case of single stars the gas infall rate onto the clouds are
$\sim 10^{-3}M_\odot~{\rm yr}^{-1}$ in our samples. The mass accretion rate
onto the star is {even} smaller than this rate, which is less
than the threshold mass accretion rate. Thus, the single stars found in
our calculations do not trace the track of the very high mass accretion rate.
{The small sample size of our simulations could lead to such an absence
of clouds with very high infall rate. But we also point out that even if we
have the case of very high mass accretion rate, the non-axisymmetry
could cause the fragmentation, which result in the formation of lower mass
stars.}
In any case, we do not find the branch of the very massive star formation
with high mass accretion rate in our samples.

We also find the stars of $\la {\rm a~few}\times M_\odot$, although they occupy
relatively small fraction.
We have to be careful to interpret this result, since we do not
have enough resolution to {study} the formation of such low mass stars.
In fact, we assume the accretion radius of $30$AU, which does not allow
the fragmentation of the disk within $30$AU. According to numerical
experiments {with} very high resolution (but very short integration time), 
the accretion disk within 30AU does fragment into many clumps, and significant fraction of them are kicked
away from the high density region via three body
interaction\citep[e.g.][]{greif12,machida_doi13}. 
Such stars seem unlikely to grow significantly, since the mass growth
rate should be smaller in less dense environments.
Considering
these theoretical evidences, the mass spectrum obtained by our simulations could
underestimate the number of low mass stars, although it is very
difficult to assess the number correctly as matters now stand.

The colors of the histogram denote the order of the birth of the stars.
It is clear that the earlier they form, the more massive they
become. We also find that the {majority of the} high mass stars of
$\ga 30M_\odot$ {are} the primary
stars, whereas the low mass stars of $\la 30M_\odot$ are not. This is
because the radiative feedback effects become significant after the
primary star grows to $20-30M_\odot$, otherwise the mass accretion is
not hindered by the radiation.

Here we mention the ``final mass'' in our simulations. The mass
accretion rate at $10^5$yr is low, but is not completely zero. The typical mass
accretion rates onto the massive stars of $\ga 100M_\odot$ are $\sim
10^{-4}-10^{-3} M_\odot{\rm yr}^{-1}$. 
In such cases the mass extrapolated
to 1Myr is a factor of two larger than the mass at $10^5$yr. 
{On the other hand}, the mass accretion rates of $10^{-4}-10^{-3} M_\odot{\rm yr}^{-1}$
 are lower than the threshold rate ($10^{-2}M_\odot{\rm yr}^{-1}$) above
 which the mass accretion could continue avoiding the
photoionization feedback\citep{hosokawa12,hirano14}. 
Thus in the present simulations, the photoionization feedback should
kick in if we have enough resolution to resolve the propagation of the ionization front. 
We have performed a very long term simulation for a minihalo that
hosts
single $170M_\odot$ star at $10^5$yr to check this conjecture. As a
result, we find the inner region of the gas is fully ionized and the
mass accretion totally stops at $\sim 6\times 10^5$yr, when the ambient
gas density becomes as low as several$\times 10^6{\rm cm^{-3}}$ and the
stellar mass is 295$M_\odot$. 
In fact, the Str\"{o}mgren radius {at} this density {($1.2\times 10^3$AU) } and the
SPH size  {($5.6\times 10^2$AU) } are comparable with
each other. This photoionization feedback should come into play at much
earlier epoch if we have enough resolution. Actually, the photoionization feedback shut off the mass accretion at $8\times 10^4$yr
in a particular grid simulation that can resolve the polar low density regions\citep{hosokawa11}. Hence, the mass
accretion seemingly stops by $\sim 10^5$yr also in our simulations if we
could resolve the propagation of ionization front.
Considering these evidences, we regard the mass at $10^5$yr as the final mass of the
stars\footnote{It is needless to say that we need RHD
simulations with high enough resolution, though.}.

\subsection{Properties of multiple stellar systems}
\begin{figure}
\plotone{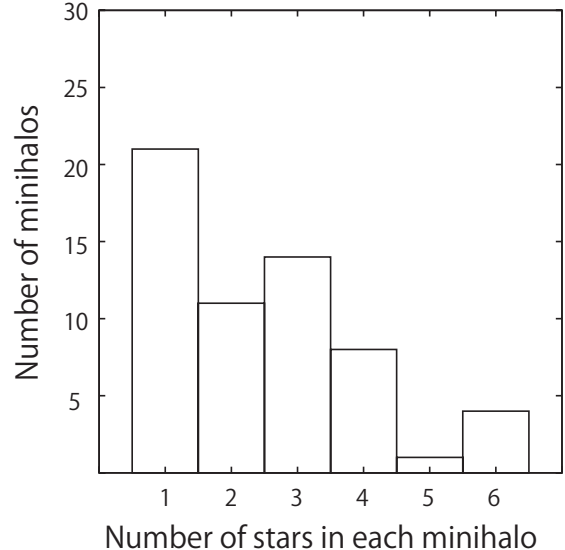}
\caption{Multiplicity of the first stellar systems are shown. Vertical
axis shows the number of minihalos while the horizontal axis is the
number of the stars in those minihalos.}
\label{fig_multi}
\end{figure}

Fig.\ref{fig_multi} shows the multiplicity of the first stellar systems
found in our numerical experiments. Approximately one-third of the 59
minihalos host single stars, while the rest of the halos have more than
two stars. {As a result, the average number of stars in a minihalo is $\sim 3$. We
also note that if we employ sink-sink merging in our simulation, the
multiplicity will decrease, although the assumption seems unlikely. 
In fact, the number of the sinks that
experience the encounter with the other sinks satisfying the merging
condition at least more than once is roughly $\sim 1/3$ of the total
number of stars. But even with such a lax condition, the average number
of stars per minihalo is $\sim 2$. }
Thus, it is clear that fragmentation of the accretion disk to form
multiple stars is a common phenomena{ on} among {forming} first
stars in various minihalos. It is
also worth noting that if we have better resolution inside
$30$AU, multiplicity will increase inevitably.

\begin{figure}
\plotone{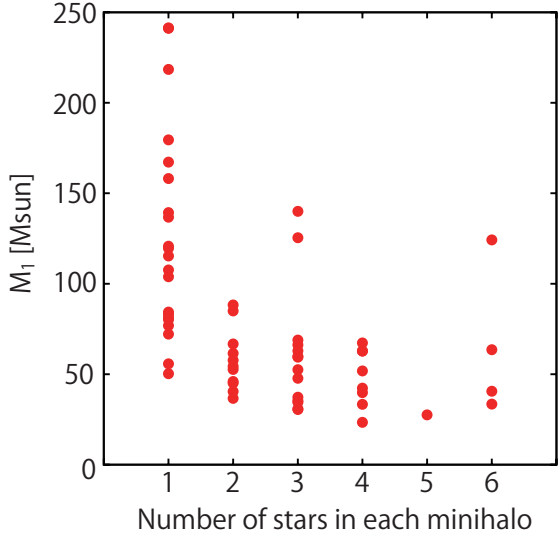}
\caption{Mass of the most massive v.s. the number of stars in a
 minihalo.}
\label{fig_most_massive}
\end{figure}

Fig.\ref{fig_most_massive} shows the mass of the primary star ($M_1$) in a
minihalo v.s. the number of stars in the minihalo. It is
clear that most of the stars with $>140M_\odot$ are born as single
stars in the present experiment, while the multiple systems scarcely contain
such {massive} stars\footnote{We find only one exception.}. 
Thus, if second generation stars form from the {ashes} of
these first stars in a single minihalo, we might be able to find the
{clear} evidence of PISN {in} the abundance pattern
of metal poor stars, which has not been found yet. We will come
back to this point in section \ref{emp_stars}.

\begin{figure}
\plotone{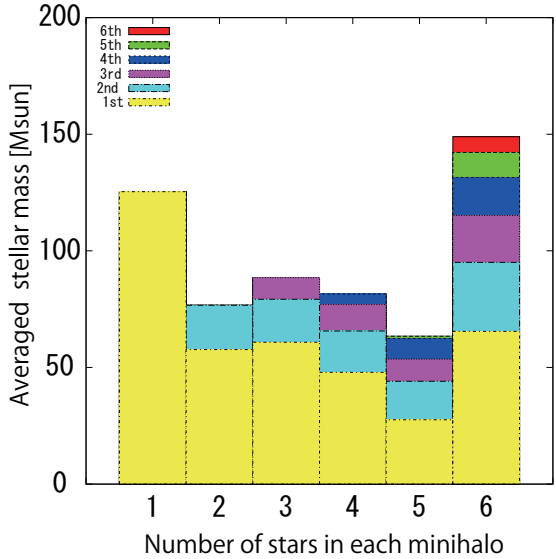}
\caption{Averaged stellar mass in a minihalo. Horizontal axis is the
number of the stars in those minihalos, while the vertical axis shows
 the stellar mass averaged over the minihalos which contain a given
 number of stars. The colors in the histogram denote to the
 fractions of the primary star (1st), secondary star (2nd), etc.}
\label{fig_budget}
\end{figure}

Fig.\ref{fig_budget} shows the budget of {the average total stellar
mass per minihalo } versus the number of stars per minihalo. The
yellow bars denote the averaged mass of the primary stars while other
colors show the mass of the 2nd, 3rd, 4th, 5th and 6th stars,
respectively (see the color legend).
It demonstrates again that very massive stars are born as single stars, and
the most massive stars in the minihalos are less than 100$M_\odot$ on
average in case they are born in multiple stellar systems. In addition,
we also find that the second massive stars are also as massive as
20-30$M_\odot$, which is massive enough to operate as the source of
ultraviolet radiation. Thus, it is crucial to include the effects
of radiative feedback from {stars other than the primaries}.

\begin{figure}
\plotone{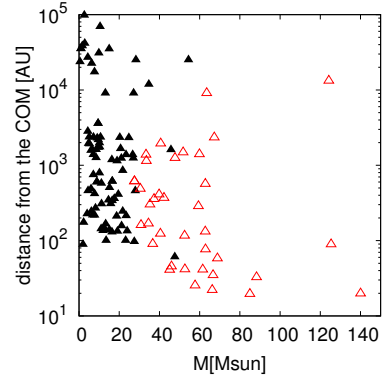}
\caption{Averaged distance from the center of mass of the stellar system to the
 stars in the minihalos. Horizontal axis is the mass of the stars, while the vertical axis shows
 the distance from the center of mass of the
 stellar system, averaged over last $2\times 10^4$yrs. Open symbols
 denote the primary stars, while the filled symbols are others. We
 omitted the stars born as single stars in this figure.}
\label{fig_sep_com}
\end{figure}

In Fig.\ref{fig_sep_com} we plot the distance from each star to the center of
mass {of the stellar system} averaged over last $2\times 10^4$yrs. Only the stars in multiple
systems are plotted. 
Open symbols denote the primary stars, while the filled symbols are the others.
As a general trend, the distances from the center of mass widely spread over 4 orders of
magnitude, i.e. form $10$AU to $10^5$AU. Secondly, the stars located
close to the center of mass tend to be massive, whereas the distant
stars are less massive, although there are some outliers. This is
because the low mass stars in multiple system tend to be kicked by
the others via three body interactions.
%Ejection'Ì‹c˜_
We also check the outward radial velocity of the relatively distant stars
of $r > 10^4$AU($\sim 0.05$pc) to assess the possibility to escape from the host minihalos.
The escape velocity of a typical minihalo is $4-5{\rm km s^{-1}}$. We
find only four of them exceed this limit. Since the total number of the 
minihalos found in the present simulations is 59, the ejection rate from the
minihalo is 0.067 star per minihalo. 
Thus the first stars born in a minihalo tend to stay
within the dark halo potential. However, as seen in
Fig.\ref{fig_sep_com}, significant fraction of stars are wandering around
$10^4-10^5$AU, which is very far from the central dense region.
The stars kicked to the distant orbit when they are still not massive
can hardly accumulate the gas{, since the surrounding gas density is low.}
{As a result, these stars cannot grow to massive stars}.  
Thus, the gravitational three body interaction is important for the mass
growth of the stars even if the
ejection rate from the minihalo is very low.

\begin{figure}
\plotone{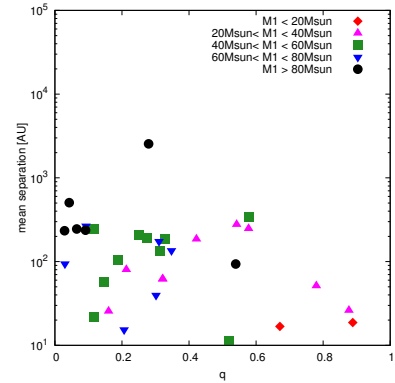}
\caption{Averaged separation of the binaries found in the simulations.
 Horizontal axis is the ratio of the mass of the two stars, while the
 vertical axis is the separation of given pairs. The distance
 is averaged over last $2\times 10^4$yrs. 
 The mass of the primary star is denoted by different symbols (see the legend).}
\label{fig_sep}
\end{figure}
We also try to find the binaries. We pick up all possible pairs of stars
in each minihalo, and assess the total internal energy of the 2-body
system as:
\begin{equation}
\epsilon =
 \frac{1}{2}\frac{m_1m_2}{m_1+m_2}\left(\bm{v}_1-\bm{v}_2\right)^2 +
 \frac{G m_1m_2}{\left|\bm{r}_1-\bm{r}_2\right|}
\end{equation}
where $m$,$\bm{v},\bm{r}$ denote the mass, velocity and position
of the stars respectively. The suffices 1 and 2 correspond to the members of
the pair. In case this energy is always negative through the last $2\times 10^4$yrs of
the simulation, we identify the pair as a binary. 
In Fig.\ref{fig_sep} the averaged separation of all binaries in the simulations are plotted against $q$ , the mass ratio of
the stars. Thus, each symbol denotes a binary in the simulation.
The mass of the primary
star of the binary is shown as the color and shape of the symbol.
The mean separation of the binary system found in our simulation is
$10-10^3$AU. We also find a marginal trend that the pairs with
more massive primary star has larger $q$.
The total number of the binaries found in the simulation is 33, so the the
frequency that a minihalo contains a binary system is $\sim 50\%$. 
%This
%fraction is lower than that was found in \citet{stacy13}, because a
%binary system can be destroyed by the encounter with other stars with
%long time integration.

\subsection{Correlation between the minihalo/cloud properties and the mass of
  the first
  stars}
\label{correlation}
\begin{figure}
\plotone{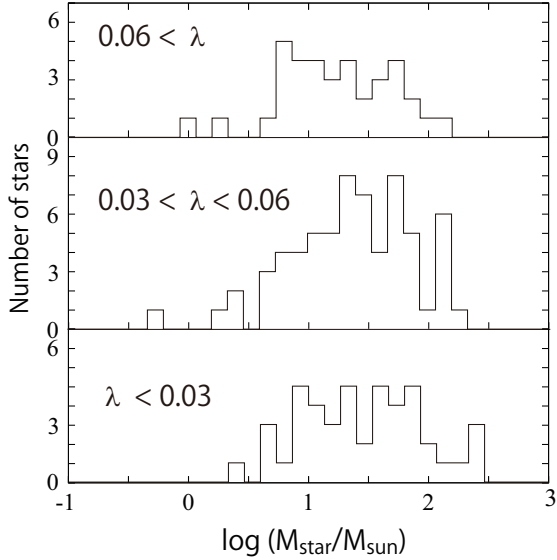}
\caption{The mass spectrum of first stars at three different
 range of $\lambda_{\rm halo}$.}
\label{fig_spectrum_lam}
\end{figure}
\begin{figure}
\plotone{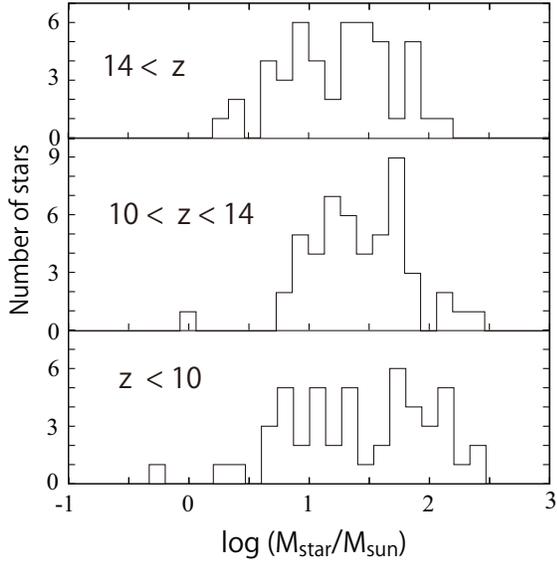}
\caption{The mass spectrum of first stars at three different
 range of formation redshift.}
\label{fig_spectrum_z}
\end{figure}
We explore the correlation between the minihalo/cloud properties and the
  mass of the first
  stars in this section. Here {a} ``cloud'' is the central dense region
  approximated as a Bonner-Ebert sphere as defined in section\ref{results_cosmological}.

Fig.\ref{fig_spectrum_lam} and \ref{fig_spectrum_z} describe the
dependence of the mass spectrum on the spin parameter and the formation
epoch of the host minihalo. We separate the minihalos into three
categories regarding the spin parameter and the
formation redshift($\lambda > 0.06,
0.03<\lambda < 0.06,\lambda < 0.03$,~~$z>14,10<z<14,z<10$) and draw the
histogram for each bin. We find no significant dependence on the spin
parameter and the formation redshift of the minihalos, although the number of stars is
not sufficient to give definitive conclusions{\citep[see also][]{hirano14,stacy14}.} This result implies that
the mass spectrum of the first stars {does not depend significantly on
the environments, which 
allows us } to use universal ``IMF'' for first stars in future cosmological simulations.
\begin{figure}
\plotone{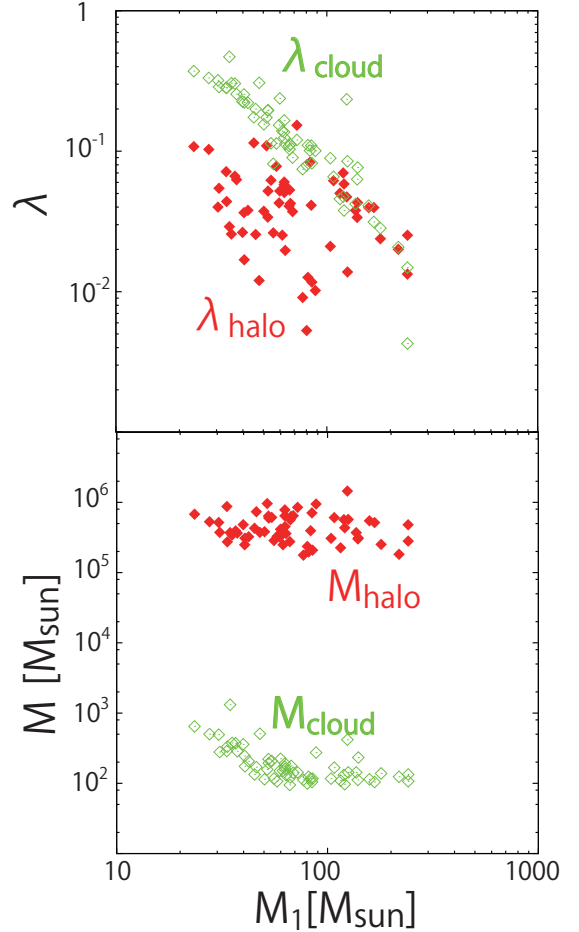}
\caption{Top: The spin parameter of
 the minihalo $\lambda_{\rm halo}$(red dots) and that of the gas cloud
 $\lambda_{\rm cloud }$(green triangles) versus the mass of the primary star. Each symbol corresponds to
 each minihalo. Bottom: Same as top panel except the spin parameters are
 replaced by mass of the minihalo and the cloud.}
\label{fig_correlation}
\end{figure}

%\begin{figure}
%\plotone{z-M1.eps}
%\caption{Formation redshifts of host minihalos versus the mass of the
% primary stars.}
%\label{fig_z-M1}
%\end{figure}

\begin{figure}
\plotone{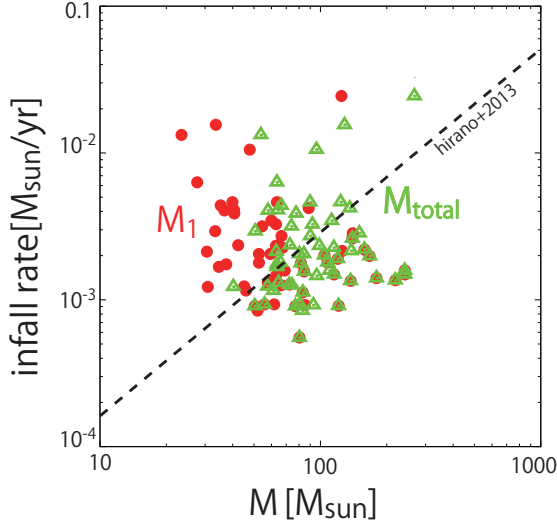}
\caption{The mass infall rate versus the mass of the primary star (red
 dots) / the total stellar mass (green triangles). Dashed
 line shows the fitted relation between the mass of the central star and
 the infall rate found in {the} 2D simulations by \citet{hirano14}.}
\label{fig_infall}
\end{figure}

Fig.\ref{fig_correlation} shows the correlation between the mass of the
primary star $M_1$ and the spin parameter of the host minihalo/cloud
(top panel), and the mass of them (bottom panel).
Each symbol corresponds to each minihalo and cloud.
First of all, we find that the mass of the primary star seems to correlate with
the spin parameter of the cloud ($\lambda_{\rm cloud}$). The relation is roughly $M_1 \propto
\lambda_{\rm cloud}^{-0.7}$. {Although it} is too complex to give analytical
justification though, it is natural qualitatively{:} higher
spin parameter inevitably results in larger disk and the gas needs to
transfer more angular momentum outward to accrete onto the primary
star. In addition, a larger disk tends to fragment to form secondaries.
In fact, the averaged $\lambda_{\rm cloud}$ over minihalos that host
single stars is $0.065$, while it is $0.20$ if averaged over the halos
that have multiple stars.
On the other hand, we do not find significant correlation between the
mass of the primary star and the spin parameter of the minihalo
($\lambda_{\rm halo}$) as expected from the results in
Fig.\ref{fig_spectrum_lam}. 
This result tells {us} that the correlation between
$\lambda_{\rm cloud}$ and $\lambda_{\rm halo}$ is weak, because the
central cloud is a much more concentrated inner system than the dark matter halo.

The bottom panel of Fig.\ref{fig_correlation} shows the correlation
between the mass of the primary star and the mass of the host
minihalo/cloud. The mass of the minihalo is basically determined by the
condition that the gas temperature is high enough to activate the H$_2$
cooling. Thus, the masses of the dark matter halos are almost identical
($M_{\rm halo} \sim 10^5-10^6\Msun$)
and seem unlikely to have correlation
with the masses of the primary stars. 
On the other hand, as for the correlation
with the cloud mass, we observe {a} weak trend that massive clouds host
low mass primary. This trend reflects the weak correlation between
$\lambda_{\rm cloud}$ and $M_{\rm cloud}$. 
%This is because the clouds with large $\lambda_{\rm
%cloud}$ require large mass for them to collapse to overcome the centrifugal
%force.
%Fig.\ref{fig_z-M1} shows the correlation between the formation redshift
%of the host minihalo and the mass of the primary star. As expected from
%the results shown in Fig.\ref{fig_spectrum_z},  we do not find clear
%trend in this plot.
{In our samples,  clouds with high spin parameters, i.e. $\lambda_{\rm cloud} \geq 0.1$, show a positive correlation between $\lambda_{\rm cloud}$ and $M_{\rm cloud}$.
However, the correlation might be misleading, since the cloud mass of
high $\lambda_{\rm cloud}$ is often overestimated with our definition of
the clouds. Except for such high-spin clouds, there is no clear
correlation between $\lambda_{\rm cloud}$ and $M_{\rm cloud}$. }

In Fig.\ref{fig_infall} we plot the mass of the
primary star (red dots) /total stellar mass (green triangles) versus the infall rate discussed in
\citet{hirano14}. Here the infall rate is defined by equation (\ref{eq:infall}).
They found that the infall rate well correlates with the mass of the
central star in their 2D RHD simulations. It is not possible 
to compare the present results directly with those in \citet{hirano14}, because their
calculation does not take into account fragmentation. However, it is
helpful to calculate the infall rate to understand the difference of two calculations.
The dashed line denotes the fit {given by} \citet{hirano14}, and red dots / green
triangles are the clouds found in our simulations. Both of the
symbols (mass of the primary and the total mass) are not well correlated
with the infall rate.
% although the trend is more or less similar to one another. 
The reason of this difference is unclear, but we guess that it mainly
originates
from the effects of fragmentation and the difference of the radiative
feedback effect caused by the fragmentation.
%We also observe that the green symbols, i.e. the plot of total stellar
%mass v.s. the infall rate is systematically below the fitted line of
%\citet{hirano14}. This tells that the accreted mass of baryons onto the
%stars is systematically larger than that in \citet{hirano14} if the halo
%properties are similar with each other. In fact, as will be discussed in
%section \ref{discussion}, our simulation could underestimate the effects of
%radiative feedback.

\subsection{Chemical imprints}
\label{emp_stars}
{The massive first stars explode as SNe and spread heavy elements
into the surrounding gas, in which the next-generation stars form. While
some of the next-generation stars also explode and contribute to further
chemical evolution of the Universe, the elemental abundance of the gas
enriched by the first stars remain in the long-lived low-mass stars or the
gas which is not recycled to ``third generation'' stars. These system
are likely to be found as metal poor systems in the
present-day Universe, such as Extremely Metal Poor (EMP) stars/ Damped Lyman$-\alpha$ (DLA) systems.
The abundance ratios of these systems preserve the nucleosynthetic
results of the first stars.
Thus, as intensively investigated in previous studies, the masses of the first
stars can be verified by comparisons between the abundance pattern of the metal-poor
systems and theoretical nucleosynthetic yields calculated from the IMF
of first stars.}

We derive theoretical yields with a stellar yield
table that includes AGB stars, core-collapse supernovae(CCSNe), and
PISNe \citep{nom13},\footnote{We assume that the
stars with $20\Msun\leq M \leq140\Msun$ explode as energetic CCSNe
with explosion energy of $\geq10^{52}$~ergs, called
hypernovae.} and the initial mass function of the first stars.
The theoretical yields are compared with the observations of two
metal-poor systems: (1) First is a metal-poor star that
formed in the early Universe and is traditionally adopted to study the
early chemical evolution \citep[\eg][]{yon13}. We adopt only normal
metal-poor stars without enhancement of C or N because
C-enhanced and N-enhanced metal-poor stars might require a specific type
of CCSNe (faint SNe, e.g., \citealt{ume02}) and self enrichment in the
low-mass metal-poor stars \citep[e.g.,][]{spi05}, respectively. (2)
Second is a metal-poor damped Lyman-$\alpha$ (DLA) system that
probes the gas-phase metal abundance at the distant Universe and is
recently studied intensively \citep[\eg][]{coo13}. Although the dust
depletion could be an observational bias, it is suggested that the dust
depletion is minimal at [Fe/H]~$<-2$ \citep[\eg][]{pet97}\footnote{Here [A/B]
$= \log_{10}(N_{\rm A}/N_{\rm B})-\log_{10} (N_{\rm A}/N_{\rm B})_\odot$,
where the subscript $\odot$ refers to the solar value and $N_{\rm A}$
and $N_{\rm B}$ are abundances of elements A and B,
respectively.}.

We test two cases of the initial mass function: the initial mass
function derived in the previous sections (IMF1) and a modified initial
mass function (IMF2), in which the mass of the first stars are reduced
by {a factor $f$.  The IMF2 is a working hypothesis to understand how
the obtained IMF(IMF1) overestimates the high mass end of the mass
spectrum,  due to the
treatment of pressure from the sink particles and the inability of
resolving the ionization front (see section \ref{IMF} and
\ref{discussion}). Here we choose 0.2 dex as $f$ for a test case.} 
The numbers of stars exploding as PISNe are 8 for
Model IMF1 and 2 for Model IMF2, {respectively}. 
First of all, the total stellar yield of all clouds is derived by
integrating the mass dependent stellar yields over the
initial mass function. The abundance pattern obtained here should be
regarded as ``a well mixed'' case, where the remnant material in a
minihalo is mixed with the others in a larger system.

\begin{figure}
\plotone{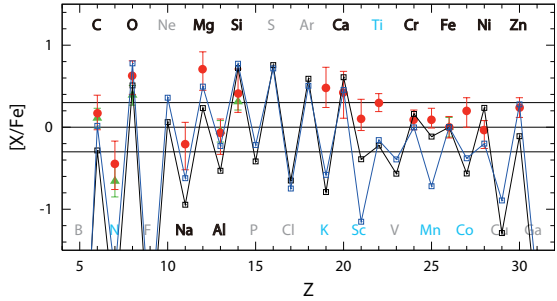}
\caption{Comparison between the abundance patterns of the total yields
 of all clouds ({\it black}: Model IMF1 and {\it blue}: Model IMF2) and
 the averaged abundance patterns of EMP stars ({\it red}:
 \citealt{cay04}) and DLA systems ({\it green}: \citealt{coo11}). }
\label{fig_allcloud}
\end{figure}

Fig.~\ref{fig_allcloud} shows comparisons between the abundance
pattern of the total yield of all clouds and the averaged abundance
patterns of the extremely metal-poor stars with [Fe/H]~$\sim-3$
(BD~--18:5550, CS~29502--042, CS~29516--024, and CS~31082--001,
\citealt{cay04})\footnote{Non-LTE effects are taken into account as
described in \cite{tom13}.} and DLA systems \citep{coo11}.
{We note that the chemical symbols indicated in gray denote the
elements without observational data. 
In addition, the theoretical yields of the elements in
cyan are known to suffer some missing ingredients in the stellar evolution/supernova model 
(rotation, mixing, neutrino, asymmetric explosion, etc., \citealt{nom13}). 
Thus we should pay attention on the elements printed as black in
the present study.
}

The abundance pattern of Model IMF1 is not in agreement with that of EMP
stars. The disagreements stem from the
large contribution of PISNe in Model IMF1 and are improved in Model
IMF2. The overall abundance pattern of the EMP stars is {mostly} reproduced
by Model IMF2 even with 2 PISNe. This demonstrates that the peculiar
chemical signatures of PISNe can be hidden by numerous CCSNe when the
number of PISNe is small and the masses of PISNe are low. 
%We note that the disagreements on
%[(Ti,Co)/Fe] is improved by explosive nucleosynthesis in the high
%entropy environment as in a jet-induced explosion \citep{tom09a}.

Compared with the averaged abundance
pattern of DLA systems, Model IMF1 and IMF2 present lower [C/Fe] and
higher [O/Fe] than the observation, respectively, and they also exhibit
over-abundance of Si due to the large contribution of massive stars with
$M\gsim50\Msun$. These results might suggest that the DLA systems are
contributed by stars with Salpeter's initial mass function or a single
Pop III CCSN as in \cite{kob11}. Indeed, the abundance ratios of an
yield of a single minihalo contributed only by CCSNe
and AGB stars is consistent with those of the DLA systems as shown
below. We note that the disagreements on [N/Fe] is improved by taking
into account the rapid rotation of the {first} stars \citep{stacy11}.

{We also consider cases with the IMFs in which the mass of the first stars are reduced
by $0.1$~dex and $0.3$~dex ($f = 10^{0.1}, 10^{0.3}$). The former case is incompatible with the
observed data as IMF1 is, whereas the latter is consistent. Thus, the
high mass end of the IMF1 seems to be overestimated more than 0.2 dex if
the abundance patterns of the low metallicity systems are mainly generated from
the nucleosynthesis of the first stars. Here after, we only consider the case of $f=10^{0.2}$ as
a representative case.}

Next, we consider the chemical enrichment in a single minihalo.
The supernova explosions are so energetic that the ejecta is not confined
in the minihalos \citep{kitayama_yoshida05,whalen08}, but the explosions
themselves could induce second generation star formation
by the compression of nearby density peaks or fragmentation of the
shocked gas \citep{greif07,sakuma09,chiaki13,dhanoa14}.
If the EMP stars formed through such process, the
abundance patterns reflect nucleosynthesis of the first stars formed
in each individual minihalo and their supernovae.

\begin{figure}
\plotone{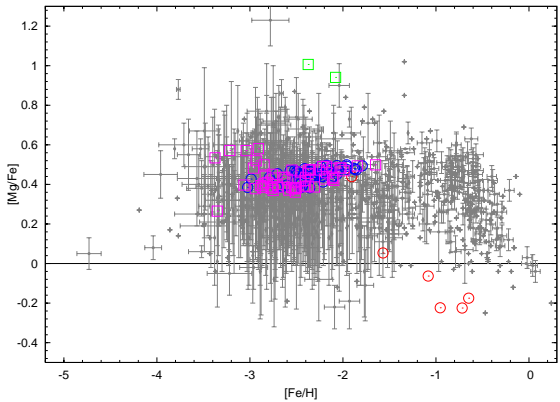}
\caption{Abundance ratios, [Mg/Fe] and [Fe/H], of the yields of
 individual minihalos enriched by PISNe ({\it red}: Model
 IMF1 and {\it green}: Model IMF2) and CCSNe and AGB stars ({\it blue}: Model
 IMF1 and {\it magenta}: Model IMF2) and the metal-poor stars ({\it
 gray}, \citealt{sud11})}
\label{fig_MgFeFeH}
\end{figure}

We simply assume that the SN yields are uniformly mixed with the gas
in the minihalo and derive the yield integrated over each halo. The
resultant [Fe/H] of the halo is derived by the ejected Fe mass and the
baryon mass of the halo.\footnote{Here, we adopt the H mass fraction
$X{\rm (H)}=0.7537$ as obtained from standard big bang nucleosynthesis
\citep{coc13}} Fig.~\ref{fig_MgFeFeH} compares [Mg/Fe] and [Fe/H] of
the yield of each halo and those of metal-poor stars collected by SAGA
database \citep{sud11}. Hereafter, LTE abundance ratios of the
metal-poor stars are adopted.
{Since most of the massive stars exploding as PISNe form as single
stars in the minihalos( Fig.\ref{fig_most_massive}), the metals of such
halos are produced only by the PISNe. 
As a result, their abundance ratio, low [Mg/Fe] in Model IMF1 and high [Mg/Fe] in
Model IMF2, are not in agreement with those of the metal-poor
stars, except for a halo with a $\sim160\Msun$ star in Model IMF1.}
\footnote{According to \citet{greif07}, $200\Msun$ PISN can
sweep $2.5\times 10^5\Msun$, which is a few times larger than the
averaged gas mass in the minihalos of our sample. Thus [Fe/H] could
sightly shift to the left especially for energetic explosions. But this
will not help to improve the disagreement.} The halos
enriched by the AGB stars and the CCSNe are distributed in the
range of $-3\lsim$~[Fe/H]~$\lsim-2$ and show [Mg/Fe]~$\sim0.4$ which is
consistent with [Mg/Fe] of the metal-poor stars.

\begin{figure*}
\plotone{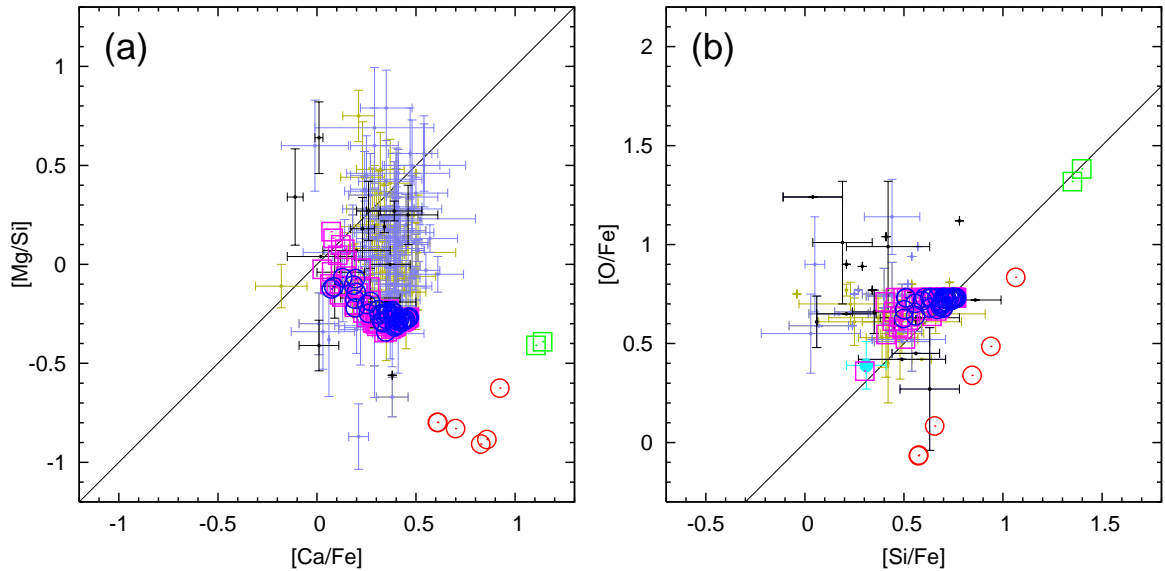}
\caption{
 Abundance ratios, (a) [Mg/Si] and [Ca/Fe] and (b) [O/Fe] and [Si/Fe], of the yields
  of individual minihalos enriched by PISNe ({\it red}: Model IMF1 and {\it green}: Model
  IMF2) and CCSNe and AGB stars ({\it blue}: Model IMF1 and {\it magenta}: Model IMF2)
  and the metal-poor stars with [Fe/H]~$<-3$ ({\it black}), $-3<$~[Fe/H]~$<-2.5$ ({\it
  light blue}), and $-2.5<$~[Fe/H]~$<-2$ ({\it yellow}) \citep{sud11}) and the average
  of the DLA systems ({\it cyan}: \citep{coo11}).
}
\label{fig_MgSiCaFe}
\end{figure*}

Figs.~\ref{fig_MgSiCaFe}(a)-\ref{fig_MgSiCaFe}(b) show the abundance
ratios between metals, [Mg/Si] and [Ca/Fe], and [O/Fe] and
[Si/Fe]. Again, the abundance ratios of the halos enriched by the AGB
stars and the CCSNe are consistent with those of the metal-poor stars,
especially with the stars with [Fe/H]~$<-3$ for the [Mg/Si] and [Ca/Fe] ratios, while
the abundance ratios of the halos enriched by the PISNe are inconsistent
with those of the metal-poor stars. It is worth {noting} that the [O/Fe] and
[Si/Fe] ratios of a single minihalo of Model IMF2 are consistent with those of
the DLA systems{ ( see overlap of magenta box and cyan filled circle in
lower-left region of panel (b) )}. 
The minihalo is enriched only by 5 less-massive CCSNe
of stars with $M\la20\Msun$ and an AGB star and also shows consistent
[C/Fe]. We also note that the diversity of abundance ratios of the halos
enriched by the AGB stars and the CCSNe are almost consistent with that
of the metal-poor stars.

The first result of chemical enrichment based on the initial mass
function of first stars suggests that the chemical imprints of PISNe
can be hidden by the larger contribution from a number of CCSNe
if the stellar yields are integrated over all halos and the
masses of the first stars are reasonably reduced by a factor of
$0.2$~dex, although the chemical signatures of PISNe cannot be hidden in
Model IMF1. This could be a solution for the longstanding problem that
the contribution of PISNe is not found in the elemental abundances of
metal-poor stars. However, we note that the chemical signatures of PISNe
should be prominent if the second-generation stars form {solely} from the gas
in the {single} minihalo, in which a PISN explodes, because such a massive star forms
as a single star. The formation of second-generation stars could be
constrained by future enhancement of the sample of metal-poor stars.

%%%%%%%%%%%%%%%%%%%%%%%%%%%%%%%%%%%%%%%%
% Dicussions 
%%%%%%%%%%%%%%%%%%%%%%%%%%%%%%%%%%%%%%%
\section{Discussions}
\label{discussion}
\subsection{Outstanding issues}
In this paper, we perform the three dimensional radiation hydrodynamics
simulations on the first star formation with long time integration to
trace the growth of the stars born in the minihalos.
In this section we describe the issues we still have to resolve as below,
in order to reach the more accurate mass spectrum of the first stars.
Interestingly, most of the ingredients we have not taken into
consideration suggest that the mass of the first stars are 
overestimated to some extent in the present simulations.
Hence, we would speculate the mass spectrum will shift to low mass
side if we include all the missing effects into the simulations.\\

{\it Resolution issues}\\

In the present numerical experiment, we still do not have enough
resolution. 
%The mass resolution of the present calculation is
%$\sim 0.5\Msun$, while the mass of the very first protostar is $\sim 10^{-3}\Msun$\citep{omukai98}.
We set the sink radius $r_{\rm sink}$ to be 30AU, but the previous
higher resolution studies of $r_{\rm sink}\la 1{\rm AU}$ (but short time
integration) revealed that
the disk within 30AU also fragments into secondary stars{ \citep[e.g.][]{clark11a,greif12}}. They also found
some of the fragments fall onto the primary, but the others are kicked to
higher orbits to survive longer time. It is not certain that these stars
survive without merging until the age of the system. However, we find
low mass stars in our simulation similar to these stars, that survive
$10^5$yrs. Thus, we can speculate that our mass spectrum seems to
underestimate the number of star at low mass side.

In addition, the simple mass accretion procedure onto the sinks employed
in this experiment also introduce an overestimation of the mass of stars. 
In the present treatment the sinks are regarded as ``black holes'', i.e.
the vicinity of the sinks are ``vacuum''. Thus the sink particles do not
push the surrounding neighbor SPH particles. As a result,  the mass
accretion rate onto the sink particles are overestimated to some extent.

We also point out that the present resolution is not enough to capture
the propagation of ionization front. 
In the SPH simulations, SPH particles are accumulated on the dense
accretion disk and thereby very few particles reside in the polar less
dense regions. As a result, the gas density at the SPH particles in the vicinity of the source
star is always comparable to the density of the disk, which is larger than $10^7{\rm cm^{-3}}$.
As was discussed in \citet{susa13}, 
the Str\"{o}mgren radius of the protostar at this density is smaller than the SPH
particle size (see Fig.9 of \citet{susa13}). Thus, the fully ionized
region does not emerge, i.e. hot ionized region of $>2\times
10^4$K does
not form even in the vicinity of the massive protostars. Consequently,
the ionization front do not break out by $10^5$yr
\footnote{The grid simulation can handle the polar low density regions more easily\citep{hosokawa11}, since the spatial resolution in such low density region is same as that in the dense disk. Lagrangian schemes like SPH change the resolution depending on the density. They have better resolution in the dense regions, but suffer low spatial resolution in less dense regions.}.
However, in the
present calculations, the H$_2$ photodissociation leads to the heating
of gas via chemical H$_2$ formation heating. As was discussed in
\citet{susa13}, H$_2$ formation heating becomes important in the
presence of photodissociative radiation, since the photodissociation
process is not a cooling process unlike the collisional dissociation processes.
In other words, after an H$_2$ molecule releases its latent heat during its
formation, photodissociation ``pumps up'' the H$_2$ molecule to
a higher
energy state (i.e. dissociated state) providing the energy by the radiation. 
Thus, strong radiative feedback exists in the present simulations even if
the photoionization feedback does not come into play. 
Thus, if the effects of photoionization are taken into consideration
properly, the radiative feedback effect will be enhanced.

We also mention another unknown factor related to the resolution. In
this calculation, the structure of the disk inside the accretion radius
is 
neglected. In fact we should have accretion disks inside the accretion radius
if we have sufficient resolution. If the disk height inside the sink
radius is not too high to shield the radiation penetrating into the outer
region of the disk ($r > r_{\rm sink}$), {the} present treatment is
{justified}. However, if the shielding by the inner disk region is significant,
our simulations overestimate the feedback effect. This is the
unknown factor that might reduce the effects of radiation and increase the mass of the stars in our
simulation, although it is not possible to infer the answer at present.
These resolution issues will be investigated in the future.\\

{\it HD cooling}\\

Importance of HD molecules has been discussed in the context of
secondary population III stars, so called PopIII.2 stars. Since the
excitation temperature is four times lower than that of the H$_2$
molecules, sufficient HD molecules cause additional cooling below $T\sim
100$K, which reduces the cloud mass/accretion rate, so does the mass of the stars.
For the formation of HD molecules, abundant H$_2$ molecules are
necessary, since the HD molecules mainly form via following reactions\citep{NU02}:
\begin{eqnarray}
{\rm D + H_2} &\rightarrow& {\rm H +HD}\nonumber\\
{\rm D^+ + H_2} &\rightarrow& {\rm H^+ +HD}\nonumber
\end{eqnarray}
In addition, it is known that the HD molecules are favored at low temperatures\citep{yoshida07,solomon73}.
%The threshold H$_2$ abundance to activate the
%HD cooling process is $y_{\rm H_2}\simeq 3\times 10^{-3}$\citep{NU02}, which is
%realized in the fossil HII regions after the death of another first
%stars or the shock heated gas in Ly-$\alpha$ cooling halos. 
Thus, {the} abundant H$_2$ at low temperature is crucial for HD cooling process,  
which is realized in the fossil HII regions after the death of other first
stars or the shock heated gas in Ly-$\alpha$ cooling halos.
\citet{hirano14} have shown that such condition could also be satisfied even
in collapsing minihalos of $10^5-10^6\Msun$.
%, although it is a relatively rare case. 
In the present simulations we do not take into account the
effects of HD cooling.  Thus, the actual mass spectrum could be shifted
to even lower mass when we include HD in our future calculations.\\

{\it Radiative feedback from other sources}\\

In the present cosmological simulations we do not include the effects of
radiative feedback from other sources, since we concentrate on the
formation of very first stars, i.e. the PopIII.1 stars. However, PopIII.2
stars also could contribute to the total star formation activities at
the cosmic dawn, we should investigate the mass spectrum of such stars.
As discussed above, the fossil HII regions could host less massive
stars on average. We also point out that the radiative feedback by a
nearby source also can reduce the mass of the stars\citep{susa09}.
Hence the mass spectrum will be shifted to lower mass side. 

The effects of {LW} background on the mass spectrum is still
uncertain\citep[e.g.][]{haiman97,machacek01,yoshida03,susa07}, although it will reduce the star formation rate definitely.\\
{The LW background intensity in unit of $10^{-21}$cgs is as
strong as 0.01 - 0.1 at $z\sim 10$\citep[e.g.][]{ahn12}, 
which could prevent the gas from cooling by destroying the H$_2$ molecules\citep{machacek01,yoshida03,ON08}.
Suppression of H$_2$ cooling heats the gas, which makes the gas cloud
gravitationally more stable. Thus, the LW background should hinder the formation of stars,
because most of the minihalos ($\sim 90\%$) form at $8 \la z \la 20$ in
our simulations. More massive halos of $\ga 10^7 M_\odot$ could collapse
due to the H$^-$ / Ly$\alpha$ cooling even in the presence of strong LW
radiation field\citep[e.g.][]{omukai08}. In such conditions, high temperature might lead
to the formation of more massive stars / massive black holes. 
Hence, the mass spectrum can potentially shifted to higher mass side if we
include the contribution from PopIII.2 stars.}
%
%{Strictly speaking, a significant fraction of the halos in our
%simulation is not PopIII.1 host, but unlikely to host any stars.
%Nevertheless, we do not find a significant correlation between the halo properties
%including the formation redshift and the mass of the stars. Hence we regard
%the hosts as PopIII.1 hosts.}
\\

{\it Streaming motion}\\

Streaming motion is the relative motion of the baryon to the dark
matter at small scale{\citep{tseli10}}. It is typically $10{\rm km~s^{-1}}$, which is larger
than the escape velocity of the minihalos that host first stars. Thus,
the minihalos that is not located at the bottom of the gravitational
potential of larger scales cannot collapse
to form stars in the presence of streaming motions.
Consequently, the formation rate of the first stars will decrease significantly\citep{naoz12}. 
On the other hand, we do not find strong correlation between the mass spectrum and
the properties of host dark matter halos such as the formation
redshifts, the spin parameters and the halo mass
(section\ref{correlation}), {which} should have correlation with the
position of the minihalos in larger scale. 
Thus, the mass spectrum might not be affected by
the streaming motion. In any case, we will explore this issue in a
forthcoming paper.\\

{\it Magnetic field}\\

Recently, interests {in} the {effect of} magnetic fields on the first star formation is
growing. The primordial gas is strongly coupled with the magnetic field
during its contraction, and the field does not dissipate from the
gas at the Jeans scale\citep{maki04,maki07}, 
{unlike in} the local interstellar gas\citep[e.g.][]{nakano_umebayashi}.
Thus, the magnetic field will be amplified more efficiently than the local
counter part. If the star forming gas cloud is magnetized with aligned
field lines of sufficient strength, magnetic breaking and jet can
transfer the angular momentum efficiently\citep{machida08}, and a single
star forms without {the} accretion disk\citep{machida_doi13}. In addition,
weaker field strength also affects the multiplicity of the
stars\citep{machida_doi13}.
However, the initial field strength, {which} can affect the dynamics of the
star formation, is larger than the expected seed field strength in cosmological
context\citep[e.g.][]{gnedin,turner,ichiki,xu,ando,doi,shiromoto}. Thus,
the magnetic field seemingly does not affect the first star formation.

On the other hand, it has been suggested that 
the magnetic field will be amplified very efficiently
by turbulent motion in the minihalo\citep[e.g.][]{schleicher10,schober12}.
If this mechanism works and the amplified small scale field inversely
cascades to larger scales smoothly, the host cloud of the first star could be
magnetized enough. However,  the
sufficient amplification to affect the dynamics of the gas has not been
shown starting from {a} very week seed field of $\sim
10^{-18}-10^{-20}$G by ab initio numerical simulations so
far\citep{sur10,federrah11,turk12,latif13}. Thus, the further progress
on this issue is desired.

\subsection{Observational signatures of the mass spectrum}
The very massive first stars are luminous, but too far to be observed directly even by next
generation huge telescopes\footnote{Their supernova explosions could be
 detected with upcoming optical and near-infrared surveys
 \citep[e.g.,][]{tanaka13}.}. 
One of the most straight-forward way to observe
the trace of the first stars is to investigate the abundance patterns of
the low metallicity systems like EMP stars. 
As was discussed in section \ref{emp_stars},
the abundance patterns of PISNe are quite different from that of CCSNe. 
If the EMP stars are born from the unmixed remnants
of the first stars, we might be able to find the clear evidence of the
existence of PISNe. 
On the other hand, we have not found the PISNe so far. 
Thus, the mass spectrum to produce many PISNe (like IMF1)
seems unlikely to represent the reality. Alternatively, if we find {any}
evidence of PISNe {in} EMP stars in the future, the frequency of the presence
of such stars would give {us the information on} the fraction of very massive stars of $>140\Msun$.
Thus, such observation would give some constraint on the high mass end
of the mass spectrum.

Another possible observation to give information on the mass spectrum of
the first stars is the hunting for the low mass zero-metallicity
stars. As was discussed in the text, recent advance {in} the first star
formation theory predicts some amounts of low mass stars of $\la
1\Msun$, which can survive the entire history of the Universe and
thereby could be found in the MilkyWay halo. The number of
such stars per single minihalo is still uncertain as of now even by
state of art numerical simulations. Therefore,
the number of such stars found by huge survey of halo stars would give
constraint on the low mass end of the mass spectrum. Even if it is not
found, we will have a strong constraint. 

\section{Summary}
We perform cosmological hydrodynamics simulations with non-equilibrium
primordial chemistry to obtain 59 minihalos that host first stars.
The obtained minihalos are used as the initial conditions of local three
dimensional radiation hydrodynamics simulations to investigate the
formation of the first stars. We employ the sink particles in the
simulations that make it possible to trace the mass growth of the stars
over $10^5$yrs. We regard the sink particles as stars and we take into
account the radiative feedback from these stars using a protostellar
evolution model. 

Then we sum up all the stars found in 59 minihalos to construct the mass
spectrum of the first stars. This is the first attempt to derive the mass
spectrum of the first stars by cosmological simulations taking into account
the fragmentation of the disk and the radiative feedback by the protostar.
As a result, the spectrum peaks at several$\times 10\Msun$, while the very
massive stars  of $>140\Msun$ also exist, which are the progenitors of PISNe.
We find 2/3 of the minihalos host multiple stellar systems whereas the
the rest of them have single stars. We also find the fraction of the
minihalos that contain binaries is $\sim 50\%$, and the mean separation of
the binaries is $10-1000$AU. 

Although most of the stars are massive ($\ga 10\Msun$), but we find a few stars are
{around} $1\Msun$. These stars are kicked by the others through
three body interactions to {a} distant less dense region, where the mass
accretion rates are very small. Since the stars of $<0.8\Msun$ can
survive through the entire history of the Universe and 
the number of these low mass stars could be enhanced if we perform
higher resolution simulations, hunting for the low mass first stars in
the local Universe {will} an important observational {attempt}.

We also investigated the chemical imprints of the mass spectrum of the
first stars on the observed low metallicity systems. We find that the
yield of the heavy elements that is obtained by integrating the mass
spectrum is not consistent with that of the low metallicity DLA systems /
EMP stars, because the fraction of PISNe is too large. If we modify the
mass spectrum to be shifted by 0.2 dex to the lower mass side, it is consistent
with the observed data, although we still have small number of PISNe.
If we consider the case that the EMP star is born in the remnant of
{an} individual minihalo, the chemical imprint of PISNe is more prominent,
because most of them are born as single stars. Thus we might be able to
find the pure abundance pattern of PISN on EMP stars if they really
formed in the ancient Universe.

%%%%%%%%%%%%%%%%%%%%%%%%%%%%%%%%%%%%%%%%
% Summary
%%%%%%%%%%%%%%%%%%%%%%%%%%%%%%%%%%%%%%%
\hspace{1cm}

\bigskip
 We thank T. Hosokawa for providing the data of protostars and S. Hirano,
 K.Omukai and N.Yoshida for fruitful discussions and careful reading of
 the manuscript. We also thank the anonymous referee for his/her
 constructive comments.
We also thank T. Takeda for his help on the instruction of the
 visualization software Zindaiji3.
We  thank the support by Ministry of
Education, Science, Sports and Culture, Grant-in-Aid for Scientific
 Research (C)22540295 (HS), (S)23224004 (NT), 
for Young Scientists (B)24740114(KH). 
KH also thanks the support by MEXT HPCI STRATEGIC PROGRAM. 
A part of the simulations were performed with the K computer at the
 RIKEN Advanced Institute for Computational Science through the HPCI
 System Research project (Project ID: hp120286, hp130026), with XC30
 "ATERUI" at the Center for Computational Astrophysics, CfCA, of
 National Astronomical Observatory , and with XE6 at the Institute for
 information Management and Communication, Kyoto University.

%%%%%%%%%%%%%%
% References 
%%%%%%%%%%%%%%

\end{document}